\documentclass[jcp,aip,preprint,floatfix,nofootinbib]{revtex4-2}
\usepackage[version=3]{mhchem} 
\usepackage{silence}
\usepackage{url}
\usepackage{graphicx}
\usepackage{dcolumn}
\usepackage{bm}
\usepackage[utf8]{inputenc}
\usepackage[T1]{fontenc}
\usepackage{mathptmx}
\usepackage{etoolbox}
\usepackage{xcolor}
\usepackage{caption}
\usepackage{subcaption}
\usepackage{float}
\usepackage{tabularx}
\usepackage{amsfonts}
\usepackage{placeins}
\usepackage{multirow}
\usepackage{geometry}
\usepackage{hyperref}
\makeatletter
\def\@email#1#2{%
 \endgroup
 \patchcmd{\titleblock@produce}
  {\frontmatter@RRAPformat}
  {\frontmatter@RRAPformat{\produce@RRAP{*#1\href{mailto:#2}{#2}}}\frontmatter@RRAPformat}
  {}{}
}%
\makeatother
\begin{document}

\title[Spin-Dep NGWFs]{Spin-Dependent Nonorthogonal Generalized Wannier Functions and their Integration with PAW and Hubbard Corrections in Linear-Scaling DFT}

\author{Miguel Escobar Azor}
\affiliation{Department of Physics, University of Warwick, Gibbet Hill Road, Coventry CV4 7AL, United Kingdom}
\affiliation{School of Chemistry and Chemical Engineering, University of Southampton, Highfield, Southampton SO17 1BJ, United Kingdom}
\author{David D. O'Regan}
\affiliation{School of Physics, CRANN Institute, and AMBER Research Centre, Trinity College Dublin, The University of Dublin, Ireland}
\author{Ali Safavi}
\affiliation{Department of Physics, University of Warwick, Gibbet Hill Road, Coventry CV4 7AL, United Kingdom}
\author{Jacek Dziedzic}
\affiliation{School of Chemistry and Chemical Engineering, University of Southampton, Highfield, Southampton SO17 1BJ, United Kingdom}
\affiliation{Faculty of Applied Physics and Mathematics, Gdańsk University of Technology, Narutowicza 11/12, 80-233 Gdańsk, Poland}
\author{Chris-Kriton Skylaris}
\affiliation{School of Chemistry and Chemical Engineering, University of Southampton, Highfield, Southampton SO17 1BJ, United Kingdom}
\author{Nicholas D. M. Hine}
\affiliation{Department of Physics, University of Warwick, Gibbet Hill Road, Coventry CV4 7AL, United Kingdom}
\email{n.d.m.hine@warwick.ac.uk}

\date{\today}

\begin{abstract}
We present a spin-dependent extension of the non-orthogonal generalized Wannier function (NGWF) formalism within the framework of linear-scaling density functional theory (LS-DFT) as implemented in the ONETEP code. In traditional LS-DFT representations, both spin channels are constrained to share a common variational basis, which limits the accuracy for systems that are spin-polarized or exhibit magnetic order. Our approach allows NGWFs to vary independently for each spin channel, enabling a more accurate representation of spin-polarization in the electronic density. We demonstrate the efficacy of this method through a series of test cases, including localized magnetic defects in two-dimensional hBN, transition metal complexes, two-dimensional van der Waals magnetic materials, 
and both bulk and nanocluster ferromagnetic Co. In each scenario, the incorporation of spin-dependent NGWFs results in enhanced accuracy for total energy calculations, improved localization of spin density, and accurate predictions of magnetic ground states. This improvement is particularly notable when combined with DFT+$U$ and DFT+$U$+$J$ corrections.
In this work, we take the opportunity to describe
the combination of DFT+$U$+$J$ and the projector-augmented wave (PAW) formalism within the 
LS-DFT framework, including how PAW participates 
in the ionic Pulay force, and in
the minimum-tracking linear response approach
for computing parameters in situ. 
Our findings demonstrate that spin-dependent NGWFs are a crucial and computationally efficient advancement in the linear-scaling DFT simulation of spin-polarized materials.
\end{abstract}

\maketitle

\section{Introduction}
Density Functional Theory (DFT) provides a computationally efficient framework for electronic structure calculations, and has become the cornerstone of first-principles materials simulation.~\cite{hohenberg1964inhomogeneous,kohn1965self} Its success lies in balancing accuracy and scalability, enabling the treatment of systems ranging from simple crystals to large biomolecules and disordered nanostructures.~\cite{jones2015density} While early DFT implementations were limited in their applicability with regards to materials with magnetic order or spin-dependent properties, the development of spin-polarized DFT extended the formalism to include spin as an explicit degree of freedom. This allowed the study of magnetic interactions, spin transport, and related phenomena.~\cite{von1972local,capelle2002nonuniqueness} Spin-polarized DFT has become a crucial tool for understanding emerging magnetic quantum materials such as 2D magnets and spintronic heterostructures,~\cite{zhang2025progress,zhao2025novel} predicting properties of actinide perovskites,~\cite{didi2025computational} and modeling spin-polarized antiferromagnets and altermagnets with exotic spin-split band structures.~\cite{guo2025spin,tamang2024newly}

Another frontier that has been opened up within DFT is that of
application to very large systems, by means of linear-scaling
DFT (LS-DFT). 
ONETEP (Order-N Electronic Total Energy Package) is a LS-DFT code that achieves plane-wave accuracy by optimizing localized nonorthogonal generalised Wannier functions (NGWFs) in situ within an
underlying, fixed psinc (plane-wave equivalent) basis.~\cite{prentice2020onetep} NGWFs are a class of localized support functions used to represent the single-particle density matrix in LS-DFT. These functions are variationally optimized during the self-consistent direct minimization process. This approach provides systematic control over accuracy while preserving sparsity, ensuring true linear-scaling behavior. This framework has also been shown to allow accurate computation of electrostatic potentials and related real-space quantities within ONETEP.~\cite{hine2011electrostatic}

Similar support-function-based strategies are employed in other LS-DFT frameworks such as BigDFT, CONQUEST, and SIESTA.~\cite{genovese2011daubechies, conquest_review, siesta_review} ONETEP has been designed from its inception to support large-scale simulations,~\cite{Hine2009_tensofthousands, Wilkinson2014_MPIOMP, dziedzic2021massively} and supports a range of features including GPU parallelism ~\cite{wilkinson2013porting,ONETEP_GPU_Docs},
advanced DFT+$U$ and constrained DFT 
methodologies~\cite{oregan_PRB_2012,PhysRevB.94.035159,prentice2020onetep,roychoudhury2018wannier}, calculation of optical absorption spectra and time-dependent response using linear-scaling formulations of TDDFT.~\cite{ratcliff2011, Zuehlsdorff2013, Zuehlsdorff2015} It incorporates a full implementation of the projector augmented-wave (PAW) method, further expanding its capability and efficiency for complex materials and strongly correlated systems.~\cite{Hine2016_PAW}

Many materials of current interest, including transition-metal complexes, magnetic 2D materials, and nanoscale systems, exhibit strong spin polarization and complex magnetic order.~\cite{Gong2017_2Dmagnetism, zhang20242d} Accurately capturing these effects requires flexibility in the treatment of spin dependence. Until now, ONETEP’s NGWF formalism employed a shared set of localized orbitals across spin channels, 
albeit with spin-dependent orbital filling and spare 
variational freedom.
This restriction constrains the NGWFs to a compromise of orbital shape and potentially limits the ability to represent spatial differences between spin-up and spin-down densities. The development of a spin-dependent NGWF framework ensures unrestricted variational freedom by allowing each spin channel to be described by its own independently optimized set of NGWFs. This enables a more faithful treatment of exchange splitting, spin localization, and spin-resolved energetics, particularly in systems with open-shell configurations and localized magnetic moments.

In this work we introduce a spin-dependent extension of the NGWF formalism in which each spin channel is described by its own independent set of localized orbitals.
The practical use of spin-dependent NGWFs is frequently in combination with two further methodologies whose implementation in ONETEP has been previously described. These are the the projector augmented-wave (PAW) method~\cite{Hine2016_PAW}, and the use of DFT+$U$ and DFT+$U$+$J$ corrections for strongly correlated systems \cite{oregan2010projector,oregan2011subspace,oregan_PRB_2012,PhysRevB.101.245137,PhysRevB.108.155141,doi:10.1021/acs.jpcc.2c04681,Sarpa_2025}. Because the use of spin-dependent NGWFs has necessitated a redesign of these functionalities, and their combination has not previously been discussed in the literature for LS-DFT, we describe here the extensions necessary to combine all three functionalities. This discussion includes the additional 
ionic Pulay force terms that arise
due to the interaction of PAW
and 
DFT+$U$~\cite{PhysRevB.102.235159}, 
and also the
(typically considerable, we observe) 
explicit PAW 
contribution to  Hubbard $U$
and Hund's $J$ parameters calculated
within the minimum-tracking linear 
response formalism. 

In Sec.~\ref{sec:spindepNGWFs} we present the theoretical and numerical foundations of the spin-dependent NGWF formalism, describe its integration with the PAW and DFT+$U$ frameworks in Secs.~\ref{sec:PAW} and \ref{sec:DFTU_PAW}, and in Sec.~\ref{sec:results} we demonstrate its performance across a diverse set of representative systems. These include point defects in hexagonal boron nitride, transition-metal molecular complexes, bilayer CrI$_3$, and large-scale cobalt nanocrystals and bulk phases. This selection spans multiple dimensionalities, from 0D molecules and defects, to 2D layered magnetic van der Waals systems, and 3D bulk and nanoscale metals, and encompasses a broad range of bonding regimes, including covalent, ionic, metallic, and non-covalent (vdW) interactions. The inclusion of open-shell and strongly correlated systems further challenges the spin representation and highlights the need for an accurate treatment of spin polarization. This diversity enables a rigorous assessment of the generality, accuracy, and robustness of the spin-dependent NGWF approach.

\section{Theory}\label{sec:theory}

\subsection{Spin-dependent NGWFs}\label{sec:spindepNGWFs}

In traditional, collinear-spin density functional theory (DFT) approaches, the single-electron density matrix for electrons of spin $\sigma$ (with $\sigma = \uparrow$ or $\downarrow$) can be expressed in terms of the Kohn–Sham orbitals $\psi_n^{\sigma}(\mathbf{r})$ as:~\cite{kohn1965self}
\begin{equation}
\rho^{\sigma}(\mathbf{r},\mathbf{r}') = 
\sum_n \psi^{\sigma}_n(\mathbf{r}) f^\sigma_n \psi^{\sigma*}_{n}(\mathbf{r}'),
\end{equation}
where $f^\sigma_n$ is the occupation number of orbital $n$ in spin channel $\sigma$. In insulating systems at zero temperature, $f^\sigma_n$ is typically $1$ or $0$ for occupied or unoccupied states, respectively, while fractional values may occur in metallic systems or at finite temperature. This formulation provides full variational freedom for the Kohn-Sham orbitals, allowing each spin channel to be represented independently, for example, through distinct sets of plane-wave coefficients.

Most approaches to linear-scaling density functional theory (LS-DFT) construct the single-electron density matrix using a localized basis of atom-centered support functions.~\cite{Goedecker1999_LSReview}
In ONETEP, these functions are referred to as nonorthogonal generalized Wannier functions (NGWFs), denoted $\phi_\alpha(\mathbf{r})$, where the index $\alpha$ labels individual NGWFs associated with different atoms. Each NGWF is strictly localized, with its spatial extent truncated beyond a spherical cutoff radius $R_\phi$, and are represented in terms of an underlying basis of psinc functions which have been well-demonstrated to provide accuracy systematically equivalent to plane-waves, at least in systems that are either not spin-polarized or have only relatively delocalized spin.~\cite{Skylaris_PRB_2002,skylaris2005introducing,Haynes_JPCS_2006} 

Within this formalism, the single-electron density matrix is represented in terms of the NGWFs and a generalized density kernel $K_\sigma^{\alpha\beta}$, which plays the role of a spin-resolved occupation matrix adapted to a nonorthogonal basis. The kernel encapsulates both the occupation and the mutual orthogonality of the Kohn-Sham orbitals. For spin-polarized systems, the density matrix for spin channel $\sigma$ takes the form:
\begin{equation}
\rho^{\sigma}(\mathbf{r},\mathbf{r}') =
\phi_{\alpha}(\mathbf{r}) K_{\sigma}^{\alpha\beta} \phi^{*}_{\beta}(\mathbf{r}'),
\end{equation}
where the presence of the spin index $\sigma$ on the kernel allows for independent population of states in each spin channel. Throughout this manuscript, we employ the summation convention that repeated Greek indices that label basis functions, support functions, projectors, or atomic sites (e.g., $\alpha$, $\beta$, $\mu$, $\nu$, etc.) are implicitly summed over unless otherwise stated, but the spin index $\sigma \in \{\uparrow, \downarrow\}$ is only summed where explicitly indicated. Labels $\alpha,\beta$ etc are subscripted for covariant quantities such as NGWFs, superscripted for contravariant quantities such as kernel matrix elements, whereas the spin index $\sigma$ is placed above or below this with no intended significance.

The ground state KS energy is defined as the minimum of a functional of the NGWFs and the density kernel:
\begin{equation}
E_{\mathrm{KS}} = \min_{\phi_\alpha(\mathbf{r}),\, K^{\alpha\beta}_\sigma} 
E[\{\phi_\alpha(\mathbf{r})\},\, K^{\alpha\beta}_\sigma]. 
\end{equation}
In the pre-existing formulation, the density kernel $K_\sigma^{\alpha\beta}$ carries a spin index, allowing for spin-resolved occupation, but the NGWFs themselves, $\phi_\alpha(\mathbf{r})$, remain spin-independent. As a result, each NGWF must simultaneously represent both spin channels, which introduces an approximation in spin-polarized systems. In practice, the main impact of this is that when constructing the total energy gradient for NGWF optimisation, we are required to average over the spin channels:
\begin{equation}
\frac{\partial E}{\partial \phi_\alpha(\mathbf{r})} = \frac{1}{2}
\sum_\sigma \mathrm{Tr} \left[ \frac{\partial E^\sigma}{\partial \hat{\rho}^\sigma }
\frac{\partial \hat{\rho}^\sigma}{\partial \phi_\alpha(\mathbf{r})}
\right].
\end{equation}
This averaged gradient is used to update the shared NGWFs $\phi_\alpha(\mathbf{r})$ during each iteration of the outer optimization loop. While this approach simplifies the optimization, it limits variational freedom, particularly in systems with strongly-broken spin symmetry.

This approximation leads to a partially restricted expansion of the Kohn-Sham orbitals, in which the same set of NGWFs is used for both spin channels:
\begin{equation}
\psi_{n}^{\sigma}(\mathbf{r}) = 
M_{n}^{\alpha \sigma} \phi_\alpha(\mathbf{r}),
\end{equation}
where $M_{n}^{\alpha \sigma}$ are the (spin-dependent) expansion coefficients relating the (non-spin-dependent) NGWFs $\phi_\alpha$ to the spin-$\sigma$ Kohn-Sham orbital $\psi_n^\sigma$. 
By contrast, full variational freedom would require allowing the NGWFs themselves to carry a spin index.
Quantifying the impact of this approximation is nontrivial, as a direct comparison requires a spin-dependent implementation of the NGWF formalism.
We therefore extend the formalism to explicitly incorporate spin dependence into both the support functions and the density kernel by assigning each NGWF a spin label, allowing the support functions, $\phi^{\sigma}_\alpha(\mathbf{r})$, to be independently optimized for each spin channel $\sigma$.~\cite{Kubler1983_spinDFT}

The one-body density matrix then takes the form:
\begin{equation}
\label{eq:dkern_spindep}
\rho^\sigma(\mathbf{r},\mathbf{r}') = 
\phi^{\sigma}_{\alpha}(\mathbf{r}) K_{\sigma}^{\alpha\beta} \phi^{\sigma*}_{\beta}(\mathbf{r}'),
\end{equation}
where $\phi^{\sigma}_{\alpha}(\mathbf{r})$ now denotes spin-dependent NGWFs
and $K_{\sigma}^{\alpha\beta}$ the spin-resolved density kernel.

With the introduction of spin-dependent NGWFs, the Kohn-Sham orbitals $\psi^\sigma_n(\mathbf{r})$ can be expanded as:
\begin{equation}
\psi^{\sigma}_n(\mathbf{r}) = 
M_n^{\alpha\sigma} \phi^{\sigma}_\alpha(\mathbf{r}),
\end{equation}
where $M_n^{\alpha\sigma}$ denotes the matrix of expansion coefficients connecting the NGWFs $\phi^{\sigma}_\alpha$ to the spin-$\sigma$ Kohn-Sham orbital $\psi^\sigma_n$ (note that the repeated $\sigma$ labels do not imply a summation). This formulation retains full variational freedom within each spin channel and ensures that the optimized NGWFs can independently adapt to spin-dependent features of the electronic structure.
In what follows, we  assume real-valued
functions for simplicity, as these
are sufficient for the treatment of
large systems (sampling at $\Gamma$) 
in the absence of
spin-orbit coupling. 

Note that the NGWF-based formalism of ONETEP has for some time been able to approximate some degree of spin-dependence in the NGWFs by employing a larger, non-minimal set of NGWFs, e.g., via the split-norm technique discussed in Appendix Section 2, of Ruiz-Serrano et al.~\cite{RuizSerrano2012_Pulay} This technique introduces artificial splitting of the NGWF norms to generate spatial 
diversity (avoidance of linear
dependence) in their initial guesses, and hence 
accommodate broken spin symmetry within a set of NGWFs. This approach
has proven very effective in benchmarking studies requiring high precision, for example in Ref.~\onlinecite{macenulty2025benchmarking}. While capable of reproducing plane-wave-level energies, this approach significantly increases the computational cost by increasing the number of overlapping pairs of NGWFs on each site. In contrast, with spin-dependent NGWFs the only contributions to the density are from spin-diagonal terms, so the current approach offers both improved physical accuracy and improved computational performance for linear-scaling DFT on magnetic systems.

\subsection{Projector Augmented Wave Formalism}
\label{sec:PAW}
The projector augmented-wave (PAW) method~\cite{blochl1994projector} enables efficient and accurate treatment of all-electron wavefunctions within a pseudopotential framework, by treating wavefunctions inside the atomic core regions on radial grids. The PAW approach is valuable for spin-polarized and strongly correlated systems as these typically involve species such as transition metals and rare-earth elements, for which norm-conserving pseudopotentials can require very high energy cutoffs. PAW works by expressing the all-electron wavefunction $|\psi^\sigma\rangle$ in terms of a smooth pseudo-wavefunction $|\tilde{\psi}^\sigma\rangle$ via a linear transformation:
\begin{equation}
|\psi^\sigma\rangle = |\tilde{\psi}^\sigma\rangle + 
\left( |\varphi_\nu\rangle - |\tilde{\varphi}_\nu \rangle \right)
\langle \tilde{p}^\nu | \tilde{\psi}^\sigma \rangle , 
\end{equation}
where \( |\varphi_\nu\rangle \) and \( |\tilde{\varphi}_\nu\rangle \) are the all-electron and pseudo partial waves, respectively, and \( |\tilde{p}^\nu\rangle \) are the PAW projector functions. This transformation allows the evaluation  of matrix elements of various operators to be performed with all-electron accuracy.

In the context of ONETEP and other linear-scaling methods using nonorthogonal support functions, this transformation implies that all inner products must be calculated with a modified overlap operator. The inner product between any two (spin-dependent) NGWFs $|\phi^\sigma_\alpha\rangle$ and $|\phi^\sigma_\beta\rangle$ is given by:
\begin{equation}
S^\sigma_{\alpha\beta} = \langle \phi^\sigma_\alpha | \hat{S} | \phi^\sigma_\beta \rangle,
\qquad
\hat{S} = 1 + 
|\tilde{p}_\mu\rangle\, O^{\mu\nu}_{\mathrm{PAW}} \langle \tilde{p}_\nu |,
\label{eq:paw_overlap_operator}
\end{equation}
where $O^{\mu\nu}_{\mathrm{PAW}}$, which is spin-independent
(the contemporary practice of PAW), 
and captures the difference between the all-electron and pseudo partial wave overlaps:
\begin{equation}
O^{\mu\nu}_{\mathrm{PAW}} = \langle \varphi_\mu | \varphi_\nu \rangle - \langle \tilde{\varphi}_\mu | \tilde{\varphi}_\nu \rangle.
\label{Eq:o_paw}
\end{equation}
Note that the matrix $O^{\mu\nu}_{\mathrm{PAW}}$ is block-diagonal in that only elements where $\mu$ and $\nu$ are on the same atomic site are nonzero.

This operator $\hat{S}$ is central to the PAW implementation in linear-scaling DFT~\cite{Hine2016_PAW} appearing in many expressions involving overlaps of pairs of functions.
Through consistent inclusion of the effect of $\hat{S}$, and the augmentation of any other required operators, it is possible to retain most of the existing functionality and algorithms in the context of PAW, as described in Ref. \onlinecite{Hine2016_PAW}, including kernel optimisation by the Haynes-Skylaris-Mostofi-Payne \cite{Haynes2008_DKopt} adaptation of LNV\cite{Li1993}, and by ensemble DFT\cite{RuizSerranoEDFT}.

In the HSMP approach, minimization of the energy is carried out with respect to the auxiliary density kernel $L^{\alpha\beta}_\sigma$, which enters the energy via the normalized kernel,
\begin{equation}
\tilde{K}^{\alpha\beta}_\sigma = \frac{N_\sigma (3LSL - 2LSLSL)_\sigma^{\alpha\beta}}{\operatorname{Tr}(S_\sigma \cdot (3LSL - 2LSLSL)^\sigma)} = \frac{N_\sigma K_\sigma^{\alpha\beta} }{\operatorname{Tr}(S^\sigma K_\sigma)}\, ,
\end{equation}
which in the second expression has been put in simpler form by defining the once-purified kernel $K_\sigma^{\alpha\beta}$ as:
\begin{equation}
{K}^{\alpha\beta}_\sigma = (3LSL - 2LSLSL)^{\alpha\beta}_\sigma \; .
\end{equation}
As we show below, this augmented overlap enters the construction of Hubbard projectors in the DFT+$U$ and DFT+$U$+$J$ frameworks.
The projectors thereby become all-electron ones, in effect, 
e.g., exhibiting oscillations in the core region. 
Their energy contributions, and derivatives with respect to matrix elements of the density kernel, the NGWFs themselves, and to the atomic position, must all be calculated accounting for this modified overlap. We therefore take this opportunity to briefly recapitulate the theory behind the use of DFT+$U$ and DFT+$U$+$J$ in the NGWF formalism, and show how it must be adapted for PAW.

\subsection{DFT+$U$ in the PAW Formalism}
\label{sec:DFTU_PAW}
The DFT+$U$ method extends 
practical approximate density functional theory (DFT) in an attempt to correct the treatment of strongly localized electrons, such as those in transition-metal $d$ or lanthanide/actinide $f$ orbitals, but
increasingly also those of other character on non-metal atoms. 
It has been observed to improve predictions of band gaps, magnetic ordering, and charge localization by penalizing fractional occupancies within a chosen correlated subspace.~\cite{anisimov1997first,PhysRevB.84.115108,PhysRevB.108.245137} 
The method pairs well with the projector augmented-wave (PAW) formalism as the latter  enables accurate all-electron treatment of semi-core orbitals, which are often influential in $d-$ and $f-$electron bandstructure.

As discussed, the PAW formalism introduces atom-centered augmentation regions to reconstruct all-electron properties, for which one set of nonorthogonal projectors is required, and likewise the DFT+$U$ method introduces Hubbard projectors $\varphi$ defining the correlated subspace, which may also be nonorthogonal~\cite{oregan2011subspace,roychoudhury2018wannier}.
These projectors are user-defined and, in general, extend spatially
well outside the PAW augmentation spheres, 
though some codes choose to define them only 
within the same spheres (a potentially severe approximation~\cite{PhysRevB.108.245137}). 
Although prior ONETEP articles have addressed DFT+$U$ and PAW separately, here we present a unified presentation of spin-resolved PAW augmentation,~\cite{Hine2016_PAW} and DFT+$U$+$J$ corrections~\cite{oregan2010projector,oregan2011subspace,oregan_PRB_2012} in the linear-scaling ONETEP framework.
All operator expectation values and overlaps involving NGWFs or Hubbard projectors must include PAW augmentation via the overlap operator $\hat{S}$. This applies to NGWF–NGWF overlaps, NGWF–projector overlaps, and all matrix elements entering energies, gradients, and forces.~\cite{blochl1994projector,kresse1999ultrasoft,Hine2016_PAW}

We first discuss the evaluation of the DFT+$U$ energy functional in this framework, then proceed to its derivatives with respect to the spin-dependent density kernel and the NGWFs, and the construction of energy-consistent Hubbard forces. The energy correction due to DFT+$U$ is expressed in terms of an occupancy matrix $n^{\sigma m'}_m = \langle \varphi_m  | \hat{\rho}^\sigma |\varphi^{m'}  \rangle$, which represents the  density matrix projected onto the Hubbard manifold. This is a block diagonal matrix, nonzero only on Hubbard sites, whose diagonal elements represent the populations of each orbital of the Hubbard site, so is expressed as a mixed tensor with one contravariant index and one covariant index~\cite{oregan2011subspace,chai_CP2K_2024}. The widely-used Hubbard energy functional of Dudarev et al.\cite{Dudarev_Hubbard_1998} can be written as:
\begin{align}
\label{eq:hubbard_energy}
E_U = & \frac{U}{2}\sum_{\sigma} \operatorname{Tr} \left[  n^\sigma - n^{\sigma} n^{\sigma}  \right],
\end{align}
where $U$ is a diagonal matrix containing the appropriate $U$ parameter for each Hubbard site (i.e., pre-defined 
subspace, typically of a given orbital character).
We define the Hubbard projection matrix $P^{\sigma}_{\alpha\beta}$ as:
\begin{equation}
P^{\sigma}_{\alpha\beta} = V^{\sigma}_{\alpha m} O_U^{mm'} W^{\sigma}_{m' \beta}. \label{eq:Proj_oc_mat}
\end{equation}
Here, $O_U^{mm'}$ is the inverse of the PAW-augmented Hubbard projector overlap matrix, hence $O_U^{mm''} \langle \varphi_{m''} |
 \hat{S} | \varphi_m' \rangle = \delta_{mm'}$, and the matrices $V^{\sigma}_{\alpha m}$ and 
its adjoint $W^{\sigma}_{m \beta}$ are the PAW-augmented overlap matrices between NGWFs and Hubbard projectors:
\begin{align}
V^{\sigma}_{\alpha m} &= \langle \phi^{\sigma}_\alpha | \varphi_m \rangle + 
\langle \phi^{\sigma}_\alpha | \tilde{p}_\mu \rangle O^{\mu\nu}_{\mathrm{PAW}} \langle \tilde{p}_\nu | \varphi_m \rangle, \label{eq:V_paw}\\
W^{\sigma}_{m' \beta} &= \langle \varphi_{m'} | \phi^{\sigma}_\beta \rangle + 
\langle \varphi_{m'} |  \tilde{p}_\mu \rangle O^{\mu\nu}_{\mathrm{PAW}} \langle \tilde{p}_\nu | \phi^{\sigma}_\beta \rangle. \label{eq:W_paw}
\end{align}
Using the definition of $\hat{\rho}^\sigma$ in terms of spin-dependent NGWFs from Eq.~\ref{eq:dkern_spindep}, we obtain the Hubbard energy expression:
\begin{align}
E_U = & \frac{1}{2} \sum_{\sigma} \operatorname{Tr} \left[ U P^{\sigma} \tilde{K}_\sigma -  U (P^{\sigma} \tilde{K}_\sigma)^2 \right],
\end{align}
which yields a corresponding term in the Hamiltonian given by
\begin{equation}
\label{eq:hubbard_ham_ngwfrep}
    H^{U,\sigma}_{\alpha\beta} = \frac{\partial E_U}{\partial \tilde{K}_\sigma^{\alpha\beta}} = \frac{1}{2} \left( U P^\sigma - 2 U P^\sigma \tilde{K}_\sigma P^\sigma \right)_{\alpha\beta} \, .
\end{equation}
Although the Hubbard projectors $\varphi_m$ are normally spin-independent by construction (taking the form of fixed pseudoatomic orbitals), spin dependence enters the projection matrix when the NGWFs are spin-dependent. In the case of self-consistent projectors constructed from NGWFs, as is optionally available in ONETEP as described in Ref.~\onlinecite{oregan2010projector}, then if the NGWFs are spin-dependent, the projectors themselves are also spin-dependent, and in fact the projected subspace is no longer shared between spin channels.

\subsubsection*{DFT+$U$ Energy gradient with respect to NGWFs with PAW}

The form of the DFT+$U$ energy gradient with respect to the density kernel is unchanged in PAW, and once the $S$, $O_U$, $V$ and $W$ matrices are appropriately augmented, the expressions presented in Appendix A of Ref.\cite{oregan_PRB_2012} can be used unchanged. However, this is not true for the the derivative of the DFT+$U$ energy with respect to a (spin-dependent) NGWF $\phi^\sigma_\alpha(\mathbf{r})$. To see this, we apply the chain rule to enumerate all dependence on $\phi^\sigma_\alpha(\mathbf{r})$:
\begin{align}
\label{eq:dE_U_dphi}
\frac{\partial E_U}{\partial \phi^\sigma_\alpha(\mathbf{r})}
= & \phantom{+}
\frac{\partial E_U}{\partial \tilde{K}^{\beta\gamma}_{\sigma}} 
\left(
\frac{\partial \tilde{K}^{\beta\gamma}_{\sigma}}{\partial K^{\delta\epsilon}_{\sigma}} \frac{\partial K^{\delta\epsilon}_{\sigma}}{\partial S_{\zeta\eta}^{\sigma}} 
+ \frac{\partial \tilde{K}^{\beta\gamma}_{\sigma}}{\partial S_{\zeta\eta}^{\sigma}}
\right)
\frac{\partial S_{\zeta\eta}^{\sigma}}{\partial \phi^\sigma_\alpha(\mathbf{r})} \notag
\\ 
& + 
\frac{\partial E_U}{\partial P^{\sigma}_{\beta\gamma}} 
\frac{\partial P^{\sigma}_{\beta\gamma}}{\partial \phi^\sigma_\alpha(\mathbf{r})}.
\end{align}
Differentiating Eq.~\ref{eq:paw_overlap_operator} with respect to 
(real-valued) $\phi^\sigma_\alpha(\mathbf{r})$ gives:
\begin{align}
\frac{\partial S^{\sigma}_{\zeta\eta}}{\partial \phi^\sigma_\alpha(\mathbf{r})}
= \phantom{+} &
\delta_{\alpha \zeta} \phi^\sigma_\eta(\mathbf{r}) + \phi^\sigma_\zeta(\mathbf{r}) \delta_{\alpha \eta}
+ \delta_{\alpha \zeta} \tilde{p}_\mu(\mathbf{r}) O^{\mu\nu}_{\mathrm{PAW}} \langle \tilde{p}_\nu | \phi^\sigma_\eta \rangle \notag \\ 
& + \langle \phi^\sigma_\zeta | \tilde{p}_\mu \rangle O^{\mu\nu}_{\mathrm{PAW}} \tilde{p}_\nu(\mathbf{r}) \delta_{\alpha \eta},
\end{align}
The augmentation-dependent terms arise from the action of the overlap operator on the PAW projectors and are captured compactly by defining $\tilde{Q}^{U,\sigma}_{\eta\zeta}$ as in Ref.~\onlinecite{Haynes2008_DKopt},
\begin{equation}
\label{eq:Qdef}
\tilde{Q}^{U,\sigma}_{\eta\zeta} = H^{U,\sigma}_{\gamma\beta} 
\left( 
\frac{\partial \tilde{K}^{\beta\gamma}_\sigma}{\partial K^{\delta\epsilon}_\sigma} \frac{K^{\delta\epsilon}_\sigma}{S^{\sigma}_{\zeta\eta}} 
+ \frac{\partial \tilde{K}^{\beta\gamma}_\sigma}{\partial S^{\sigma}_{\zeta\eta}}
\right).
\end{equation}
Next, the second term of Eq.~\ref{eq:dE_U_dphi} requires the derivatives of the projected occupation matrix with respect to the NGWFs. For clarity we adjust the indices of the projected occupation matrix defined in Eq.~\ref{eq:Proj_oc_mat} for spin $\sigma$ to read:
\begin{equation}
\label{eq:Proj_oc_mat_2}
P^{\sigma}_{\beta\gamma} = V^{\sigma}_{\beta m} O_U^{mm'} W^{\sigma}_{m'\gamma}\, .
\end{equation}
Then, by combining Eq.~\ref{eq:Proj_oc_mat_2} with Eq.~\ref{eq:V_paw} and Eq.~\ref{eq:W_paw} we can write the derivative of $P^{\sigma}_{\beta\gamma}$ with respect to $\phi^\sigma_\alpha(\mathbf{r})$ as:
\begin{align}
\label{eq:dP_dphi}
\frac{\partial P^{\sigma}_{\beta\gamma}}{\partial \phi^\sigma_\alpha(\mathbf{r})}
= \quad
& \delta_{\alpha \beta}  \left( \varphi_m(\mathbf{r})\, O_U^{mm'} W^{\sigma}_{m'\gamma} + \tilde{p}_\mu(\mathbf{r})\, O^{\mu\nu}_{\mathrm{PAW}} \langle \tilde{p}_\nu | \varphi_m \rangle O_U^{mm'} W^{\sigma}_{m'\gamma} \right)  \notag \\
  + & \left(V^{\sigma}_{\beta m} O_U^{mm'}\, \varphi_{m'}(\mathbf{r})\, + \, V^{\sigma}_{\beta m} O_U^{mm'} \langle \varphi_{m'} | \tilde{p}_\nu \rangle O^{\nu\mu}_{\mathrm{PAW}} \, \tilde{p}_\mu(\mathbf{r}) \right) \delta_{\alpha \gamma} \, .
\end{align}
Using Eq.~\ref{eq:hubbard_ham_ngwfrep} the dependence of the energy on the projection matrix can be compactly expressed as:
\begin{equation}
\label{eq:dE_U_dP}
\frac{\partial E_U}{\partial P^{\sigma}_{\beta\gamma}} = K^{\gamma\delta}_\sigma H^{U,\sigma}_{\delta\epsilon} P_{\sigma}^{\epsilon\beta} \, ,
\end{equation}
where $P_\sigma^{\epsilon\beta} = (P_\sigma^{-1})^{\epsilon\beta}$ is the contravariant inverse of the projection matrix, which we do not have in explicit form. However, it can be avoided as since $P^{-1}.(VO_U W) = 1$, we can write
\begin{equation}
\label{eq:H_U_simplification}
H_{\delta\epsilon}^{U,\sigma} P_\sigma^{\epsilon\beta}V_{\beta m}^\sigma O_U^{mm'} = V_{\delta m} H^{m m'}_{U,\sigma} \, ,
\end{equation}
where $H^{m m'}_{U,\sigma}$ is the DFT+$U$ term in the  Hamiltonian, expressed in the subspace of the projectors, defined by $H_{\alpha\beta}^{U,\sigma} = V_{\alpha m}^\sigma H^{mm'}_{U,\sigma} W_{m'\beta}^\sigma$. 
Finally, inserting Eqs.~\ref{eq:Qdef}, \ref{eq:dP_dphi}, \ref{eq:dE_U_dP} and \ref{eq:H_U_simplification} into Eq.~\ref{eq:dE_U_dphi}, resolving all the Kronecker deltas, noting that $W_{m'\gamma}=V^{\dagger}_{\gamma m'}$, and collecting coefficients of each function we obtain:
\begin{align}
\label{eq:final_dftu_grad}
\frac{\partial E_U}{\partial \phi^\sigma_\alpha(\mathbf{r})}
&= 
2 K^{\alpha\delta}_\sigma V_{\beta m}^\sigma H_{U,\sigma}^{mm'} \varphi_{m'}(\mathbf{r}) \notag \\
&\quad + 2 K^{\alpha\delta}_\sigma V_{\beta m}^\sigma H_{U,\sigma}^{mm'}  \langle \varphi_{m'} | \tilde{p}_\nu \rangle O^{\nu\mu}_{\mathrm{PAW}} \tilde{p}_\mu(\mathbf{r}) \notag \\
&\quad \quad + 2 \tilde{Q}^{U,\sigma}_{\alpha\zeta} \phi^\sigma_\zeta(\mathbf{r}) + 2 \tilde{Q}^{U,\sigma}_{\alpha\zeta} \langle \phi^\sigma_\zeta | \tilde{p}_\mu \rangle O^{\mu\nu}_{\mathrm{PAW}} \tilde{p}_\nu(\mathbf{r}) \, .
\end{align}
This expression contains four distinct contributions: the first two terms arise from the direct and augmention terms of the overlap between NGWFs and Hubbard projectors. The third and fourth terms arise from the direct and augmentation terms of the correction that arises from the change in the derivative of the overlap matrix as the NGWFs change. These latter terms share a common form with many
non-DFT+$U$ terms in the NGWF gradient, and they are included in the gradient 
automatically, in practice, by,  including $H^{U,\sigma}$ in the full Hamiltonian in the NGWF representation, which is used to construct a full $\tilde{Q}$
for the full Hamiltonian. As discussed above, Hubbard projectors could become spin-dependent if spin-dependent NGWFs are used as self-consistent projectors, but in standard usage they are spin-independent PAOs so the $\sigma$ labels are dropped.

\subsubsection*{DFT+$U$ Forces}
We now derive the DFT+$U$ contribution to atomic forces in the presence of PAW augmentation. This builds upon expressions previously presented in the nonorthogonal projector formalism of Ref.~\onlinecite{oregan2010projector}, the DFT+$U$ energy derivatives in nonorthogonal basis from Ref.~\onlinecite{oregan_PRB_2012}, and the PAW implementation in ONETEP presented in Ref.~\onlinecite{Hine2016_PAW}. Our formulation here extends these to the spin-dependent case, ensuring full consistency of energies and forces in linear-scaling DFT+$U$ with PAW.

To compute the force on atom $I$ due to the DFT+$U$ correction, we begin from the chain rule expression for the total derivative of the Hubbard energy with respect to ionic positions, noting that the forces are evaluated with the total energy already having been variationally minimised with respect to $\{K^{\alpha\beta}_\sigma\}$ and $\{\phi^\sigma_\alpha(\mathbf{r})\}$ so no derivative terms arise from these:
\begin{equation}
\label{eq:dftU_force}
F_U^I = -\frac{\partial E_U}{\partial \mathbf{R}_I}
= -\sum_\sigma \frac{\partial E_U}{\partial P^{\sigma}_{\alpha\beta}} \cdot \frac{\partial P^{\sigma}_{\alpha\beta}}{\partial \mathbf{R}_I}.
\end{equation}

The first term was already encountered in the NGWF gradient, while the second term encapsulates the change in the projected occupation matrix as the atoms move. The projection $P^{\sigma}_{\alpha\beta}$ as defined in Eq.~\ref{eq:Proj_oc_mat}, $V^{\sigma}_{\alpha m}$ and $W^{\sigma}_{m \beta}$ from Eqs.~\ref{eq:V_paw} and  ~\ref{eq:W_paw} respectively combine to give:
\begin{equation}
\label{eq:dP_dR_raw}
\frac{\partial P^{\sigma}_{\alpha\beta}}{\partial \mathbf{R}_I} =
\frac{\partial V^{\sigma}_{\alpha m}}{\partial \mathbf{R}_I} \, O_U^{mm'} \, W^{\sigma}_{m'\beta}
+ V^{\sigma}_{\alpha m} \, O_U^{mm'} \, \frac{\partial W^{\sigma}_{m'\beta}}{\partial \mathbf{R}_I}.
\end{equation}
where $\partial O_U^{mm'}/\partial \mathbf{R}_I = 0$ for displacements
that do not change the augmented 
Hubbard projector overlap matrix.
For different treatments of cases where $\partial O_U^{mm'}/\partial \mathbf{R}_I \ne 0$,
we refer the reader to Ref.~\onlinecite{roychoudhury2018wannier,
PhysRevB.102.235159}.

Next, noting that the
atom-position derivative of the matrix 
$O_\mathrm{PAW}^{\mu\nu}$ vanishes
as it is an atom-centered block-diagonal 
object, we may expand the
derivative of the NGWF-Hubbard projector
overlap matrix as:
\begin{align}
\label{eq:dV_dR_raw}
\frac{\partial V^{\sigma}_{\alpha m}}{\partial \mathbf{R}_I} = &
\Big\langle \phi^\sigma_\alpha \Big| \frac{\partial \varphi_{m}}{\partial\mathbf{R}_I} \Big\rangle \delta_{J(m),I} \notag \\
& +
\Big\langle \phi^\sigma_\alpha \Big| \frac{\partial \tilde{p}_\mu}{\partial\mathbf{R}_I} \Big\rangle O_\mathrm{PAW}^{\mu\nu} \Big\langle \tilde{p}_\nu \Big| \varphi_m \Big\rangle \delta_{J(\mu),I} \notag \\
& +
\Big\langle \phi^\sigma_\alpha \Big| \tilde{p}_\mu \Big\rangle O_\mathrm{PAW}^{\mu\nu} \Big\langle \frac{\partial \tilde{p}_\nu}{\partial\mathbf{R}_I} \Big| \varphi_m \Big\rangle \delta_{J(\nu),I} \notag \\
& + 
\Big\langle \phi^\sigma_\alpha \Big| \tilde{p}_\mu \Big\rangle O_\mathrm{PAW}^{\mu\nu} \Big\langle \tilde{p}_\nu \Big|  \frac{\partial \varphi_m}{\partial\mathbf{R}_I}\Big\rangle \delta_{J(m),I} \; , 
\end{align}
and similarly for its adjoint, as 
\begin{align}
\label{eq:dW_dR_raw}
\frac{\partial W^{\sigma}_{m' \beta}}{\partial \mathbf{R}_I} = 
&
\Big\langle \frac{\partial \varphi_{m'}}{\partial\mathbf{R}_I}  \Big| \phi^\sigma_\beta\Big\rangle \delta_{J(m'),I} \notag \\
& +
\Big\langle \varphi_{m'} \Big| \frac{\partial \tilde{p}_\mu}{\partial\mathbf{R}_I} \Big\rangle O_\mathrm{PAW}^{\mu\nu} \Big\langle \tilde{p}_\nu \Big| \phi^\sigma_\beta \Big\rangle \delta_{J(\mu),I} \notag \\
& + 
\Big\langle \varphi_{m'} \Big| \tilde{p}_\mu \Big\rangle O_\mathrm{PAW}^{\mu\nu} \Big\langle \frac{\partial \tilde{p}_\nu}{\partial\mathbf{R}_I} \Big| \phi^\sigma_\beta \Big\rangle \delta_{J(\nu),I} \notag \\
& + 
\Big\langle  \frac{\partial \varphi_{m'}}{\partial\mathbf{R}_I} \Big| \tilde{p}_\mu \Big\rangle O_\mathrm{PAW}^{\mu\nu} \Big\langle \tilde{p}_\nu \Big| \phi^\sigma_\beta\Big\rangle \delta_{J(m'),I} \; .
\end{align}

We make use of the fact that the gradient of an atom-centered function with respect to movement of the atom center, at a given point, is the negative of the gradient of the function at that point. Including augmentation terms in $V$ and $W$ then allows us to define the force response matrix $\mathbf{X}^{\sigma}_{\alpha\beta}$ to include augmentation:
\begin{align}
\mathbf{X}^{\sigma}_{\alpha m} = &
 \langle \phi^{\sigma}_\alpha | \nabla \varphi_m \rangle 
 + \langle \phi_\alpha | \nabla \tilde{p}_\mu \rangle O_\mathrm{PAW}^{\mu\nu} \langle \tilde{p}_\nu |   \varphi_m \rangle \notag \\
& + \langle \phi^{\sigma}_\alpha | \tilde{p}_\mu \rangle \, O^{\mu\nu}_{\mathrm{PAW}} \langle  \nabla \tilde{p}_\nu | \varphi_m \rangle 
+ \langle \phi^{\sigma}_\alpha |  \tilde{p}_\mu \rangle \, O^{\mu\nu}_{\mathrm{PAW}} \langle  \tilde{p}_\nu | \nabla \varphi_m \rangle  
\, ,\label{eq:XJ_paw}
\end{align}
We can thus write:
\begin{equation}
\label{eq:dP_dR_Xdef}
\frac{\partial P^{\sigma}_{\alpha\beta}}{\partial \mathbf{R}_I} =
-\mathbf{X}^{\sigma}_{\alpha m} O_U^{mm'} W^\sigma_{m'\beta}
-  V^\sigma_{\alpha m'} O_U^{m'm} \mathbf{X}^{\sigma\dagger}_{m\beta}
\end{equation}
where for each atom $I$ the partial sum over $m$ runs only over projectors on that atom.
We then substitute Eqs.~\ref{eq:dE_U_dP}, ~\ref{eq:H_U_simplification} and \ref{eq:dP_dR_Xdef} into Eq.~\ref{eq:dftU_force} to arrive at the final expression for the DFT+$U$ contribution to the force:
\begin{align}
F_U^{I,\sigma} &= 
-2 \sum_\sigma \mathrm{Re} \left\{ 
\tilde{K}^{\alpha\beta}_\sigma V^\sigma_{\beta m} H_{U,\sigma}^{mm'} \mathbf{X}^{\sigma\dagger}_{m'\alpha } \,
\right\},
\label{eq:finalU_force} 
\end{align}
where as above for each atom $I$ we sum only over the $m$, $m'$ associated with that atom.

\begin{figure}[ht]
\centering
\includegraphics[width=0.45\columnwidth]{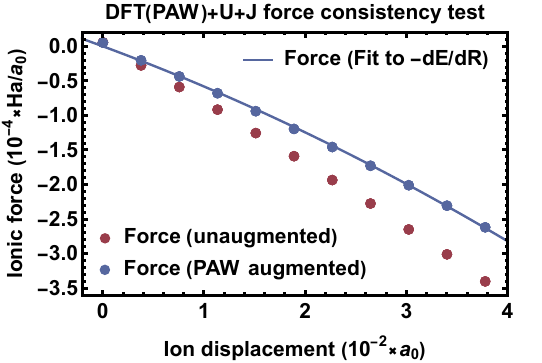}
\caption{A  test for the
consistency between 
DFT+$U$+$J$ total energy gradients,
with respect to ionic displacement, 
and DFT+$U$+$J$ forces including
PAW terms, as defined by Eqns.~\ref{eq:finalU_force},\ref{eq:finalJ_force}, and \ref{eq:tildeX}. 
The curve is the analytical derivative
of a least-squares fit of a two-parameter
(quadratic and cubic) polynomial, including
energy data points outside of the range shown. Shown are data points for 
single force components (in the direction
of displacement) on the Fe atom
in the high-spin [Fe(NCH)$_6$]$^{2+}$  complex, 
both with (PAW augmented) and
without (unagumented) the second term
of Eq.~\ref{eq:tildeX}.
}
\label{fig:Pulay}
\end{figure}

At the time of writing, we have implemented
an approximated form of PAW augmentation
in DFT+$U$ in which the overlap 
of PAW projectors and Hubbard projectors
on different atoms is neglected.
Thus, we apply 
$\langle \varphi_m \rvert 
\tilde{p}_\mu \rangle \approx 
\langle \varphi_m \rvert 
\tilde{p}_\mu \rangle 
 \delta_{I(m),J(\nu)}  $, and 
 as this feeds into the construction
 of $O_U^{m m'}$, no loss of 
 effective Hubbard projector  
 orthonormalization is created. 
While lifting this approximation
is desirable and planned, 
it seems a reasonable one
for localized projectors and typical
inter-atomic distances. It is likely 
 a mild approximation, on balance, 
 in comparison
 with the dominant paradigm of
 DFT(PAW)+$U$, i.e., that of using
 $\lvert \tilde{p}_\mu \rangle $ 
 in place of 
 $\lvert \varphi_m \rangle $.
 Under this approximation, 
 notwithstanding, the
 force response matrix reduces
 to its first two terms only
 \begin{align}
\mathbf{\tilde{X}}^{\sigma}_{\alpha m} = &
 \langle \phi^{\sigma}_\alpha | \nabla \varphi_m \rangle 
 + \langle \phi_\alpha | \nabla \tilde{p}_\mu \rangle O_\mathrm{PAW}^{\mu\nu} \langle \tilde{p}_\nu |   \varphi_m \rangle , 
 \label{eq:tildeX}
 \end{align}
since the latter two involve the
gradient of the overlap
$\langle \varphi_m \rvert 
\tilde{p}_\mu \rangle  $
of now comoving-only functions.
The numerical validity of this
expression is illustrated in
Fig.~\ref{fig:Pulay}, 
which was generated by performing
total-energy calculations (without
restarts) with small displacements to 
the central Fe atom in the high-spin
state of the complex 
[Fe(NCH)$_6$]$^{2+}$. Calculation
were performed with placeholder 
non-zero $U$ values on subspaces of all atoms, and non-zero $J$ on Fe, 
in a cubic simulation cell of side length
20~\AA{}, but otherwise as described
in Ref.~\onlinecite{macenulty2025benchmarking} with
ionic geometry from Ref.~\onlinecite{Mariano2020}.

\subsection{Combining DFT+$U$+$J$ with PAW}

The DFT+$U$+$J$ method extends the DFT+$U$ framework by introducing a Hund's exchange parameter $J$, enabling a more 
comprehensive treatment of 
spin-dependent correlation effects.~\cite{liechtenstein1995density,PhysRevB.84.115108, Moore2024_UJ}  The formalism accounts not only for intra-spin electron-electron repulsion, but also explicitly incorporates inter-spin exchange interactions, potentially 
enhancing the accuracy of predictions 
for magnetic ordering and 
electronic localization in
correlated-electron materials.
This picture is not without its complications, 
however, particularly for the total 
energy~\cite{PhysRevB.107.L121115,macenulty2025benchmarking}.
Here, we focus on the  simplified 
rotationally invariant 
DFT+$U$+$J$ functional of 
Ref.~\onlinecite{PhysRevB.84.115108}.

The energy correction is defined in terms of the spin-resolved projected site occupations as:
\begin{align}
E_{U+J} = & E_U + E_J \\ 
= & \sum_\sigma \operatorname{Tr}\left[ \frac{U-J}{2}  (n^\sigma -n^\sigma n^\sigma) \right] + \sum_\sigma \operatorname{Tr}\left[\frac{J}{2} n^\sigma n^{\bar{\sigma}} \right] \\
=& 
\sum_{\sigma}   \, \mathrm{Tr} \left[ \frac{U - J}{2} \left(P^\sigma K_\sigma - (P^\sigma K_\sigma)^2 \right) \right] \notag
\\ & +   \sum_\sigma \mathrm{Tr}\left[ \frac{J}{2} P^\sigma K_\sigma P^{\bar{\sigma}} K_{\bar{\sigma}} \right], \label{eq:energy_u+j}
\end{align}
where $\bar{\sigma}$ represents the opposite spin to $\sigma$, and like $U$ in Eq.~\ref{eq:hubbard_energy}, $J$ here is a diagonal matrix holding $J$ values for each correlated site. The second term introduces coupling between the spin channels and enforces parallel spin alignment consistent with Hund's rule. 

The contribution from $E_J$ to the Hamiltonian for spin channel $\sigma$ is:
\begin{equation}
H^{J, \sigma}_{\alpha\beta} = \frac{\partial E_{J}}{\partial K^{\beta\alpha}_\sigma} = J P^{\sigma} K_{\bar{\sigma}} P^{\bar{\sigma}}
\, .
\end{equation}
The equivalent of Eq.~\ref{eq:dE_U_dP} is:
\begin{equation}
\label{eq:dE_J_dP}
\frac{\partial E_J}{\partial P^{\sigma}_{\beta\gamma}} = K^{\gamma\delta}_\sigma H^{J,\sigma}_{\delta\epsilon} P_{\sigma}^{\epsilon\beta} \, ,
\end{equation}
so we can calculate all gradients arising from the $J$-dependent energy.
The contribution from $E_J$ to the NGWF gradient has the same form as Eq.~\ref{eq:final_dftu_grad}, with $H_{J,\sigma}^{mm'}$ substituted for $H_{U,\sigma}^{mm'}$, and the total force (once again substituting in the projector response matrix $\mathbf{X}^{\sigma}_{m\beta}$) gains a term
\begin{align}
F_J^I = & -\frac{\partial E_J}{\partial \mathbf{R}_I}
= -\sum_\sigma \frac{\partial E_J}{\partial P^{\sigma}_{\alpha\beta}} \cdot \frac{\partial P^{\sigma}_{\alpha\beta}}{\partial \mathbf{R}_I} \\
=& -2 \sum_\sigma \mathrm{Re} \left\{ 
\tilde{K}^{\alpha\beta}_\sigma V^\sigma_{\beta m} H_{J,\sigma}^{mm'} \mathbf{X}^{\sigma\dagger}_{m'\alpha } \,
\right\},
\label{eq:finalJ_force} 
\end{align}
where $H_{J_\sigma}^{mm'}$ is the Hamiltonian corresponding to the $J$-term in the Hubbard projector basis.

The same PAW-augmented definitions of overlaps and projection matrices $P^{\sigma}$ are used, ensuring that the force contributions from the $J$ term remain consistent with the PAW formalism and the spin-polarized NGWF representation.

\subsection{Combining Minimum-Tracking
Linear Response for $U$ 
and $J$ with PAW}

In this work, we will use the
minimum-tracking 
variant~\cite{Linscott2018_UJdet, Moynihan2017_scfU}
of the widely used finite-difference
linear-response formalism 
for calculating the Hubbard $U$
and Hund's $J$ parameters 
from first-principles. 
This has previously been used in 
studies with ONETEP, including
Refs.~\cite{PhysRevB.101.245137,PhysRevB.108.155141,doi:10.1021/acs.jpcc.2c04681,Sarpa_2025,macenulty2025benchmarking}, 
and it has recently been 
implemented also
in CP2K~\cite{chai_CP2K_2024}.
The formalism side-steps
the need for matrix inversions
or sampling the DFT+$U$ matrices
anywhere but at the ground state
(given a finite perturbation). 
For the case of $U$, it 
calculates the rate
of change of the subspace-and-spin
averaged internal 
potential, with respect
to the subspace occupancy.
That is, when 
the most energy-efficient 
(per constrained DFT arguments)
form of potential is used to vary that
occupancy, namely
$\hat{v}_{\mathrm{ext}} = \alpha \hat{S} \lvert \varphi^m  \rangle \langle \varphi_m \rvert \hat{S} $. The formula for the Hubbard $U$ for a given 
subspace may be written as
(recalling that paired 
opposing indices
 are summed over)
\begin{align}
U = \frac{1}{2} \frac{d \left( \langle \varphi_m \rvert \left( \hat{v}^\uparrow_{\mathrm{KS}} 
+\hat{v}^\downarrow_{\mathrm{KS}} \right) \lvert \varphi^m \rangle - 2 \alpha \right) }{ d \left( n_{m'}^{\uparrow m'}
+ n_{m''}^{\downarrow m''}\right)
\langle \varphi_{m'''} \rvert
\varphi^{m'''} \rangle} ,
\end{align}
where both the fully-converged ground-state Kohn-Sham potential
$\hat{v}^\sigma_{\mathrm{KS}}$ 
(technically a pseudo-potential, 
in the PAW sense)
and fully-converged ground-state occupancy matrix are simultaneously 
parameterized by
the external perturbation strength $\alpha$.
The $ - 2 \alpha$ excludes the
artificial external perturbation
from the definition of $U$, 
as only the variation
of the internal potential 
(including all screening effects
of unconstrained degrees of freedom, 
e.g., from outside the subspace)
is relevant.
The final inner product in the
denominator should simply be 
an integer, e.g., 5 for $3d$ orbitals.

The corresponding formula for $J$, 
which quantifies subspace spin magnetization-magnetization self-interaction~\cite{PhysRevB.84.115108,Linscott2018_UJdet,lambert2023use}, 
which may be associated
with subspace static correlation error~\cite{PhysRevB.107.L121115,10.1063/1.5091563}, is
\begin{align}
J = - \frac{1}{2} \frac{d \left( \langle \varphi_m \rvert \left( \hat{v}^\uparrow_{\mathrm{KS}} 
-\hat{v}^\downarrow_{\mathrm{KS}} \right) \lvert \varphi^m \rangle - 2 \beta \right) }{ d \left( n_{m'}^{\uparrow m'}
- n_{m''}^{\downarrow m''}\right)
\langle \varphi_{m'''} \rvert
\varphi^{m'''} \rangle},
\end{align}
where each quantity is now instead
parameterized by a perturbation 
strength
$\beta$ in 
the energetically optimal potential
for varying subspace magnetization, 
 $\hat{v}^{\uparrow}_{\mathrm{ext}} 
=\beta \hat{S} \lvert \varphi^m \rangle  \langle \varphi_m \rvert \hat{S} = 
 - \hat{v}^{\downarrow}_{\mathrm{ext}}$.

These potentials, perturbations, and
occupancies are all augmented in 
PAW, i.e., they include
core electron contributions
and the Hubbard projectors are 
rendered, implicitly, all electron
ones including oscillations in the
core region. 
Within the PAW construction, 
furthermore, 
the potential acting upon the
valence pseudo-electrons includes
a contribution that, approximately
speaking, is an augmentation of
the potential by the PAW transformation
within the core regions. 
For details, see Ref.~\onlinecite{Hine2016_PAW}, 
but suffice to say
that it may be written in the
form $\hat{v}_{\mathrm{PAW}}^\sigma
= \lvert \tilde{p}_\mu \rangle 
D^{\sigma \mu \nu}_{\mathrm{PAW}}
\langle  \tilde{p}_\nu \rvert$.
This term must not be neglected
in $\hat{v}_{\mathrm{KS}}^\sigma$
when calculating $U$ or $J$ and, 
as we show below, this
contribution may make up the 
dominant part of those parameters.

\section{Calculation Details}\label{sec:results}
All following linear-scaling DFT calculations are performed using modified versions of release 7.0 of the ONETEP code\cite{prentice2020onetep}, all relevant functionality of which has been incorporated into the latest release. Corresponding plane-wave DFT calculations for benchmarking purposes are performed with Quantum ESPRESSO~\cite{Giannozzi_2017} (QE). Unless otherwise specified, all ONETEP and QE calculations were performed using the PBE exchange-correlation functional~\cite{Perdew1996PBE} and JTH-PBE pseudopotentials.~\cite{Jollet2014JTH} The atomic structures and input files were generated using the Atomic Simulation Environment (ASE),~\cite{Larsen2017ASE} which facilitated consistent setup and cross-code compatibility.

Where appropriate, we can visualize the enhancement in spin polarization arising from spin-dependent NGWFs, by computing the local spin density:
\begin{equation}
\rho_s(\mathbf{r}) = \rho_\uparrow(\mathbf{r}) - \rho_\downarrow(\mathbf{r}),
\label{eq:spin_den}
\end{equation}
where $\rho_s(\mathbf{r})$ denotes the spin-density field.~\cite{Kubler1983_spinDFT} To isolate the effect of spin-channel adaptivity in the NGWF representation, we define a difference field:
\begin{equation}
\Delta \rho_s(\mathbf{r}) = \rho_s^{\mathrm{spin-dependent}}(\mathbf{r}) - \rho_s^{\mathrm{non-spin-dependent}}(\mathbf{r}),
\label{eq:spin_den_diff}
\end{equation}
which highlights regions where variational freedom in the spin-dependent NGWFs improves the representation of spin symmetry breaking. 

To produce spin-density maps, the three-dimensional difference field was projected onto the $xy$ plane by summation over the third Cartesian coordinate:
\begin{equation}
\Delta \rho_s^{\mathrm{proj}}(x, y) =
\sum_z \Delta \rho_s(x, y, z),
\label{eq:integrated_spin_den_diff}
\end{equation}
yielding planar maps that illustrate the spatial localization and selectivity of spin-polarized variational enhancement. These projections are plotted on a consistent colour scale in e/bohr$^2$, with axes in bohr, to facilitate visual comparison across systems.

\section{Results}

In this section, we systematically evaluate the performance and physical significance of spin-dependent NGWFs in ONETEP. We selected four representative categories of systems, with each subsection focusing on a representative physical context in which spin-dependent NGWFs offer a clear advantage over conventional non-spin-dependent formulations. These span different dimensionalities, magnetic phenomena, and degrees of electronic delocalization. Our test set includes: (i) localized defect states in insulating 2D materials, (ii) spin-state energetics in transition-metal complexes, (iii) magnetic ordering in 2D van der Waals materials, and (iv) metallic ferromagnetism in bulk and nanostructures. This diverse selection allows us to explore how spin-adaptive variational freedom affects accuracy and convergence in different spin-polarized regimes, and to examine the roles of spin-adaptive NGWFs, and of PAW-augmented DFT+$U$(+$J$), in capturing spin-broken symmetry, magnetic energetics, and localized orbital features across distinct electronic regimes.

For local defects in 2D insulators and for transition metal complexes, we assess how spin-dependent NGWFs improve the accuracy of spin-resolved electronic levels and the energy differences between competing spin states, namely, midgap states in defective hBN and high-spin versus low-spin configurations in ligand-field-split Fe(III) complexes. For 2D magnets such as CrI$_3$ we investigate the delicate energy balance between ferromagnetic and antiferromagnetic stacking configurations, and how spin-channel flexibility improves predictive power in meV scale ordering. For metallic systems, including bulk Co and Co nanocrystals, we focus on spin density distribution, density of states, and total energy stabilization arising from spin-adaptive variational freedom.

In each case, results from spin-dependent NGWF calculations are compared to non-spin-dependent ONETEP runs and, where available, to benchmark plane-wave DFT calculations (Quantum ESPRESSO) and hybrid-functional results (ORCA). This comparative analysis enables us to assess where the impact of spin-dependent NGWFs and Hubbard corrections is most significant, and where the different available formalisms can be made to more precisely agree with each other.

\subsection{Localized Defects in Hexagonal Boron Nitride}
We first examine the performance of spin-dependent NGWFs in describing localized, spin-polarized states induced by point defects in two-dimensional materials. As a model system, we consider monolayer hexagonal boron nitride (hBN), a wide-bandgap insulator known for its chemical stability, atomically smooth surfaces, and compatibility with van der Waals heterostructures,~\cite{dean2010boron, watanabe2004direct} in which context it is generally regarded as electronically inert.\cite{Magorrian_TMD_hBN_2022} Although pristine hBN is nonmagnetic, substitutional impurities can significantly alter its electronic and magnetic properties by introducing midgap states that are both spatially localized and spin-polarized due to unpaired electrons.~\cite{Attaccalite2011hbn}

We focus on two well-studied point defects involving carbon substitution: a carbon atom replacing a boron atom (C$_B$) and one replacing a nitrogen atom (C$_N$). Both break the sublattice symmetry of the hBN lattice and produce midgap states with distinct spin and energetic features. The C$_B$ defect introduces a singly occupied state near the conduction band minimum (CBM), while the C$_N$ defect leads to an unoccupied state close to the valence band maximum (VBM). Examining the energetics of these spin-polarized states tests whether the variational flexibility of localized support functions helps with the description of spatial distribution of spin density around magnetic defects. Recent studies have emphasized the structural and optoelectronic complexity of such carbon-related centers, underscoring the need for accurate \textit{ab initio} methods to characterize them.~\cite{koperski2020midgap, wu2024ab}
%

Structural relaxations for both defect configurations were performed using Quantum ESPRESSO, and the resulting geometries served as the input for ONETEP single-point calculations, which were repeated in both spin-dependent and non-spin-dependent NGWF formulations.

\begin{figure}[ht]
\centering
\includegraphics[width=0.95\columnwidth]{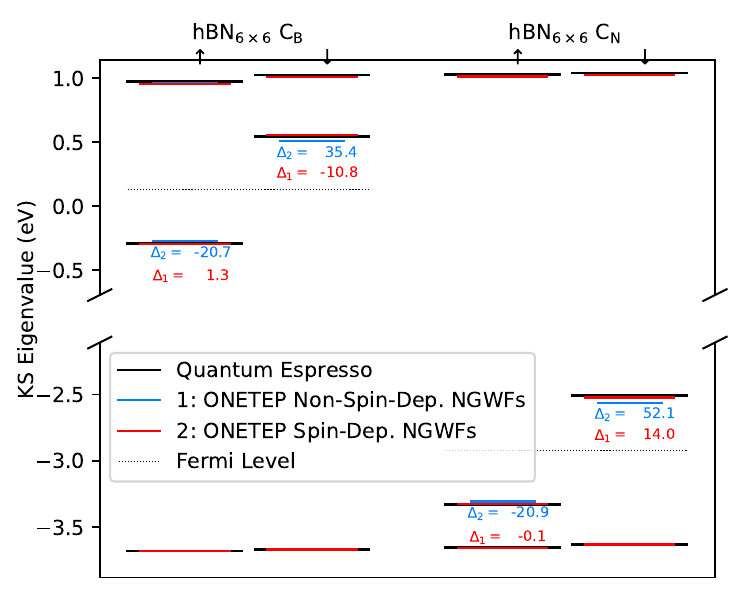}
\caption{Defect levels (in eV) for substitutional carbon defects in a $6 \times 6$ hBN supercell. Horizontal lines indicate the positions of spin-up and spin-down defect levels computed using plane-wave DFT in Quantum ESPRESSO (green), traditional non-spin-depenent NGWFs (blue), and spin-dependent NGWFs (red). The $\Delta$ values quantify deviations of ONETEP results from the plane-wave reference, and are quoted in meV.}
\label{fig:defect_hBN}
\end{figure}

Figure~\ref{fig:defect_hBN} compares the computed KS eigenstates associated with the defects, for both configurations using spin-dependent NGWFs (red), non-spin-dependent NGWFs (blue), and a reference plane-wave calculation performed with Quantum ESPRESSO (green). All energy levels are shown on an absolute scale, and spin-up/down components are resolved where applicable. The plane-wave results, serving as a benchmark, correctly reproduce the aforementioned spin-resolved defect physics.

While the non-spin-dependent NGWFs already exhibit qualitative agreement with the PW results, with deviations between 20 and 55~meV, the spin-dependent NGWFs yield excellent numerical agreement with the plane-wave reference, reducing discrepancies in defect level positions to the few-meV level. For the C$_\mathrm{B}$ defect, the occupied midgap level is well-separated from the CBM and exhibits clear spin polarization. Spin-dependent NGWFs reproduce this spin splitting with deviations under 11~meV, capturing both the energetic position and spin-resolved character of the defect state. In contrast, non-spin-dependent NGWFs introduce larger errors, approaching 40~meV in opposite directions for spin-up and spin-down levels. A similar pattern is observed for the C$_\mathrm{N}$ defect, which produces an unoccupied level just above the VBM. Spin-dependent NGWFs result in deviations of +0.09~meV and –20.88~meV for spin-up and spin-down channels, respectively, compared to the plane-wave reference. Non-spin-dependent NGWFs again show greater errors: +13.96~meV and +52.13~meV. These comparisons  highlight the limitations of a shared set of NGWFs in resolving localized spin-polarized states near the CBM and the need for spin-channel flexibility when treating spin-polarized defect states near band edges.

The improvements from the spin-dependent formalism arise from removing the constraint that both spin channels share the same set of NGWFs. By allowing spin-up and spin-down densities to be represented with separate sets of NGWFs, there is more scope to optimize the energy with respect to the charge and spin distributions around defects. This added flexibility leads to more accurate defect-level energies and improved resolution of spin polarization, both of which are critical for describing midgap states localized around defects.

\subsection{Transition Metal Complexes}

We next present results on a series of Fe(III) coordination clusters with varying number of Fe ions and varying ligand environments, focusing on both high-spin and low-spin states to assess spin-state energetics.~\cite{Gutlich2004SpinCrossover} These range from mononuclear to trinuclear iron(III) complexes. The mononuclear structures were generated using Avogadro\cite{Hanwell_Avogadro_2012,Avogadro} and the dinuclear structure\cite{Mathur_Fe2Structure_2002} and trinuclear structure\cite{Georgopoulou_Fe3Structure_2010} were taken from the Cambridge structural database\cite{Groom_CSD_2016} and processed using Mercury (version 3.8).\cite{Macrae_Mercury_2006,Macrae_Mercury_2020} Geometry optimizations were performed using ORCA (version 5.0.2, with single-point calculations using ORCA 5.0.3).~\cite{Neese2020ORCA,NeeseORCAUpdate2022} The PBE0 hybrid functional~\cite{Perdew1996PBE,PerdewExactExchangeDFT1996,adamo1999} and the def2-TZVP basis set\cite{Weigend2005} with an auxiliary basis of def2/J.\cite{Weigend2006} were used in calculations. Alongside the interest in the improved description from spin polarized NGWFs, another key question for this system is the extent to which DFT+$U$ calculations can be used as an accurate substitute for much more costly hybrid functional calculations.
We therefore perform both DFT and DFT+$U$ calculations with ONETEP and Quantum ESPRESSO for comparison, though note that due to the different level of theory, exact comparability of energetics is not expected in these cases. For example, even though they are both using pseudoatomic orbitals as DFT+$U$ projectors, there is still not exact equivalence of the projector functions between ONETEP and QE, as discussed more extensively in Section \ref{sec:twodimvdW}.
The ORCA PBE0 calculations are included here because hybrid DFT methods are generally expected to provide reliable levels of accuracy for spin-state energetics in transition metal complexes~\cite{harvey2004dft, Pierloot2017}, particularly combined with the high degree of convergence obtainable with a triple-zeta def2-TZVP basis set. The inclusion of a fixed fraction of exact (Fock) exchange in PBE0 would be expected to partially correct self-interaction errors, which should improve the description of the energy differences between spin states.

System-specific convergence criteria were applied, with NGWF radii and kinetic energy cutoffs converged individually for each system.
For calculations with ONETEP and QE, the optimized gas-phase structures were embedded in cubic supercells with at least 10~\AA{} of vacuum to mitigate interactions between periodic images.
For the plane-wave DFT and DFT+$U$ calculations carried out using Quantum ESPRESSO, we used equivalent energy cutoffs and pseudopotentials to the corresponding ONETEP calculations, and fine FFT grids with a density cutoff of four times the plane-wave cutoff.
Calculations in both ONETEP and Quantum ESPRESSO performed using PBE+$U$ both used $U = 4$~eV, unless
stated as having been directly 
calculated.

\begin{figure}[h]
\centering
\begin{subfigure}{.5\columnwidth}
  \centering
  \includegraphics[width=\linewidth]{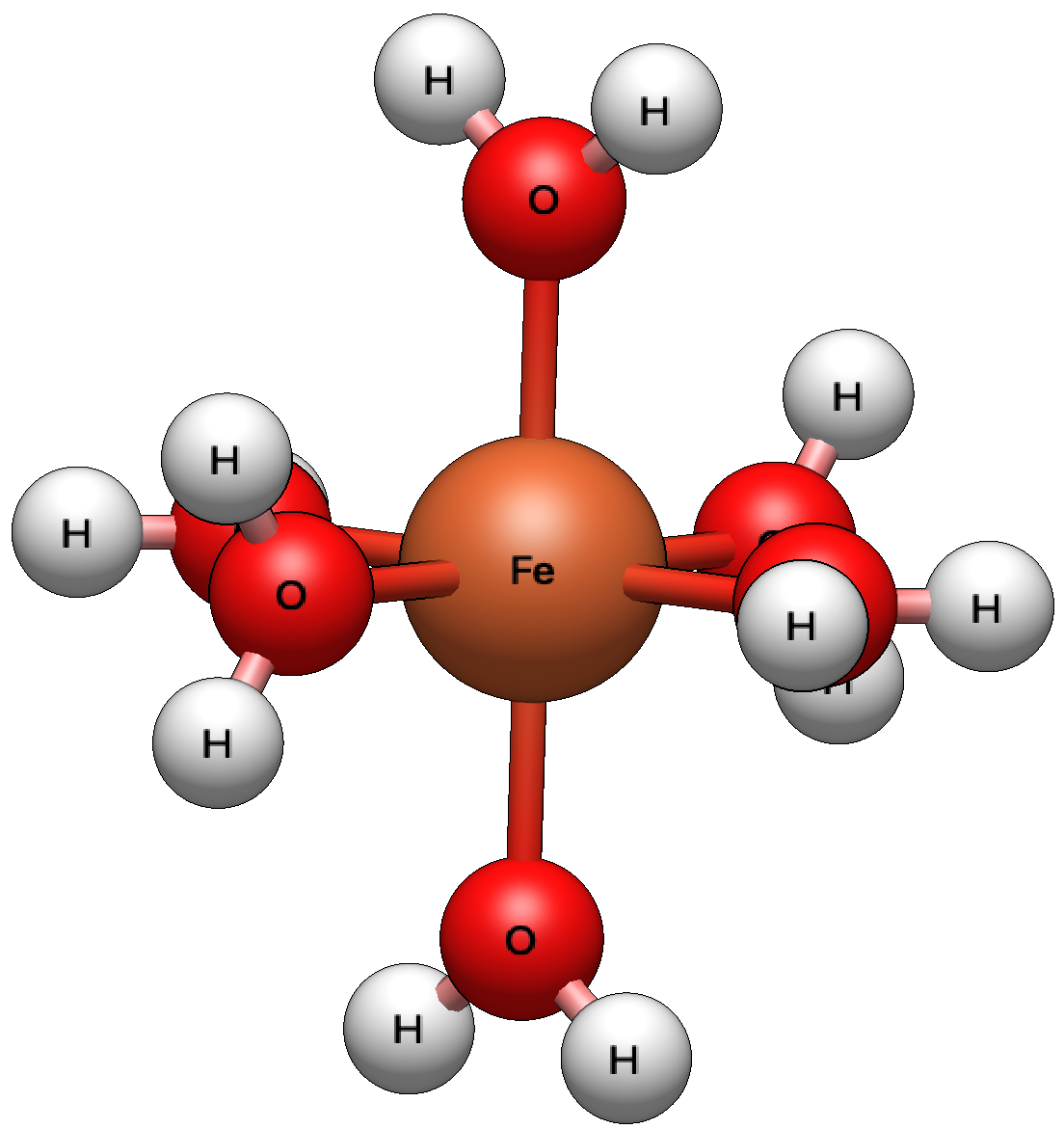}
  \caption{$\mathrm{[Fe(H_2O)_6]^{3+}}$}
\end{subfigure}%
\begin{subfigure}{.5\columnwidth}
  \centering
  \includegraphics[width=\linewidth]{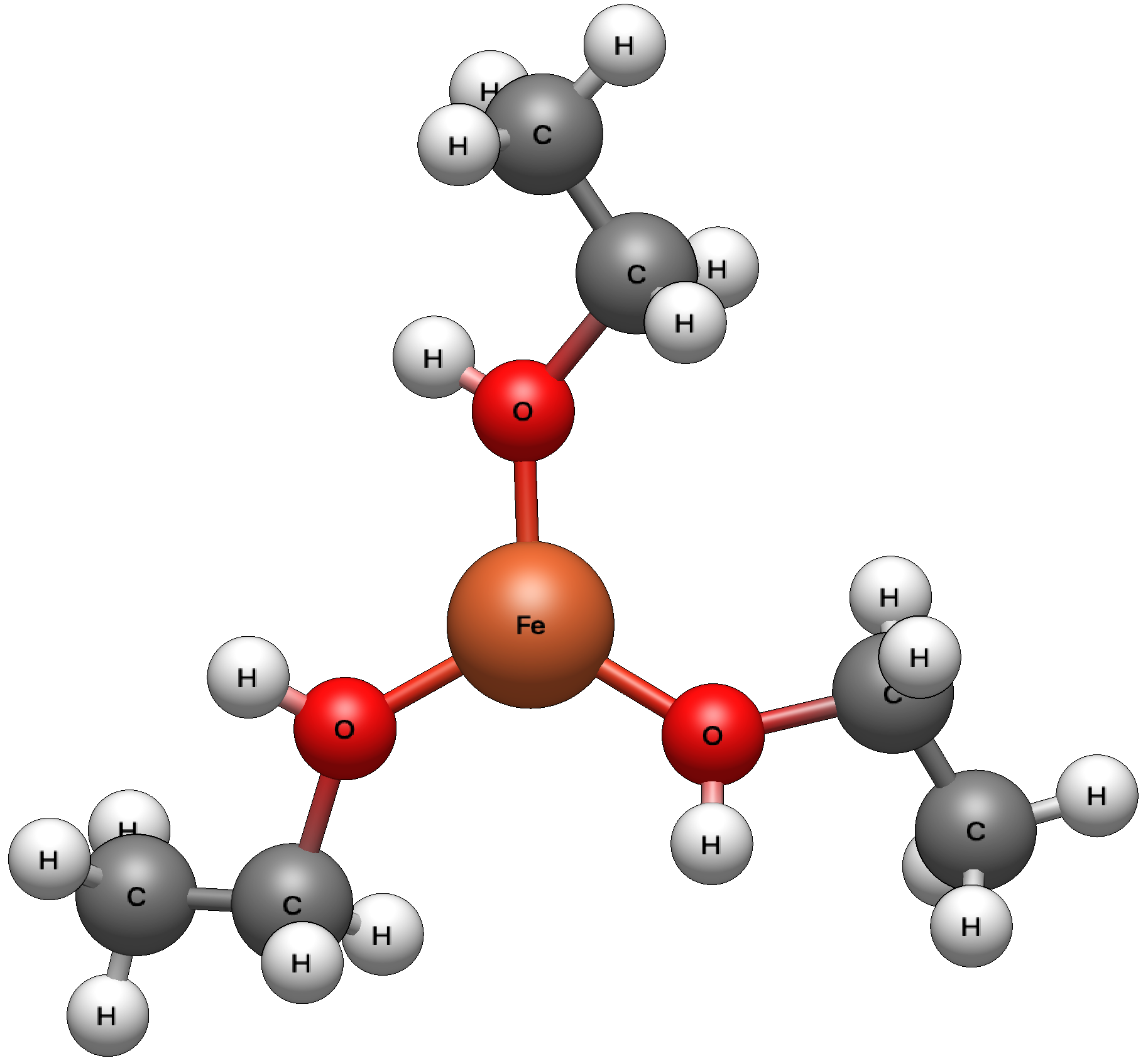}
  \caption{$\mathrm{[Fe(EtOH)_3]^{3+}}$}
\end{subfigure}
\begin{subfigure}{.5\columnwidth}
  \centering
  \includegraphics[width=\linewidth]{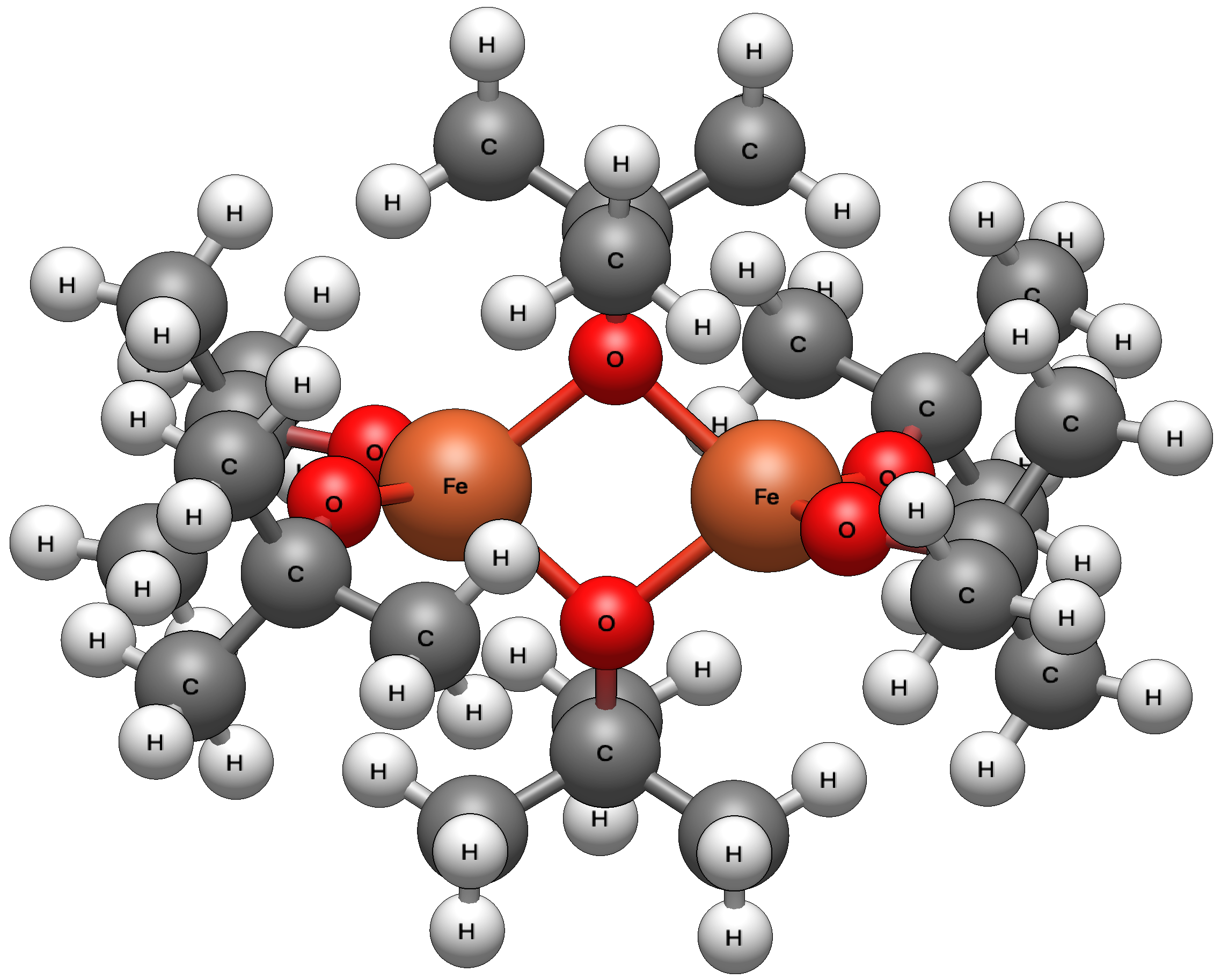}
  \caption{$\mathrm{[Fe_2(OtBu)_6]}$}
\end{subfigure}%
\begin{subfigure}{.5\columnwidth}
  \centering
  \includegraphics[width=\linewidth]{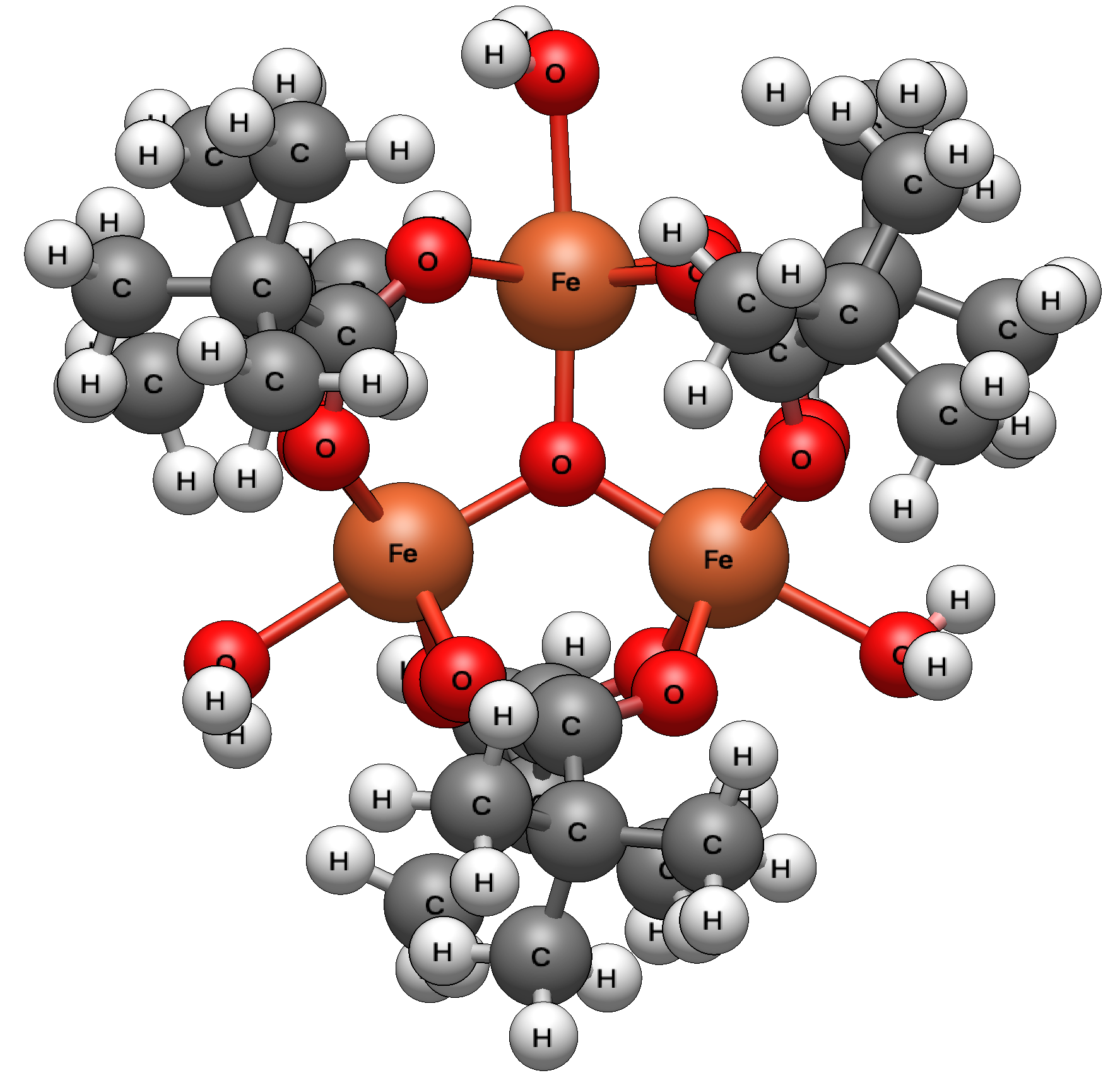}
  \caption{$\mathrm{[Fe_3O(Piv)_6(H_2O)_3]^+}$}
\end{subfigure}
\caption{Structures of the optimized iron(III) clusters in the high-spin state, based on geometries optimized using the PBE0 functional.} 
\label{iron structures}
\end{figure}

The optimized structures of the transition metal clusters studied are visualized using Avogadro 2 (version 1.99.0) in Figure~\ref{iron structures}.\cite{Hanwell_Avogadro_2012} For these fixed optimized structures we calculate the total energy differences between the high-spin (HS) and low-spin (LS) states, defined as $\Delta E_{\text{HS} \rightarrow \text{LS}} = E_{\text{HS}} - E_{\text{LS}}$. These are listed in Table~\ref{tab:HS_LS_energies}. In ONETEP, results are shown for both non-spin-dependent and spin-dependent NGWFs, for PBE and PBE+$U$. These are compared with equivalent energy differences, firstly from Quantum ESPRESSO PBE and PBE+$U$ calculations, and secondly from ORCA PBE0 hybrid functional calculations. 

\FloatBarrier

\begin{table*}[ht]
    \centering
    \caption{
        High-spin to low-spin single-point energy differences 
        (in eV) for selected Fe(III) complexes, computed using PBE NO $U$.
        a) ONETEP with non-spin-polarized NGWFs;
        b) ONETEP with spin-polarized NGWFs;
        c) Quantum ESPRESSO;
        d) ORCA with PBE0/def2-TZVP.
    }
    \begin{tabular}{cccccc}
        \hline
        \textbf{Complex} & \textbf{Method} & \textbf{(a) ONETEP NSD} & \textbf{(b) ONETEP SD} & \textbf{(c) Quantum ESPRESSO}  &\textbf{(d) ORCA} \\
        \hline \multirow{2}{*}{
        $\mathrm{[Fe(H_2O)_6]^{3+}}$}  & PBE          & -1.29 & -1.35 &  -1.37 & -1.06 \\
                                       & PBE+$U$/PBE0   & -1.86 & -1.91 & -1.67  & -1.69 \\
        \hline \multirow{2}{*}{
        $\mathrm{[Fe(EtOH)_3]^{3+}}$}  & PBE           & -1.63 & -1.71 &  -1.72 & -1.24 \\
                                       & PBE+$U$/PBE0     & -1.23 & -1.25 & -1.75  & -1.84 \\
        \hline \multirow{2}{*}{
        $\mathrm{[Fe_2(OtBu)_6]}$}     & PBE           & -2.74 & -2.83 &  -0.81 & -2.14 \\
                                       & PBE+$U$/PBE0     & -3.55 & -3.63 & -2.61  & -3.74 \\
        \hline \multirow{2}{*}{
        $\mathrm{[Fe_3O(Piv)_6(H_2O)_3]^+}$} & PBE       & 0.32  & 0.18  &  -0.42 & 0.74 \\
                                             & PBE+$U$/PBE0 & -2.36 & -2.49 & -2.32  & -2.56 \\
        \hline
    \end{tabular}
    \label{tab:HS_LS_energies}
\end{table*}

The use of spin-polarized NGWFs in ONETEP (Table~\ref{tab:HS_LS_energies}, column b) leads to a small but consistent increase in the calculated spin-state energy differences $\Delta E_{\text{HS} \rightarrow \text{LS}}$, relative to non-spin-polarized NGWFs (column a), corresponding to greater stabilization of the high spin state. This occurs in both DFT and DFT+$U$, and in the former brings the result into better agreement with plane-wave DFT results, at least in the mononuclear cases.
For the mononuclear complexes $\mathrm{[Fe(H_2O)_6]^{3+}}$ and $\mathrm{[Fe(EtOH)_3]^{3+}}$, spin-polarized NGWFs result in increases of 0.05~eV and 0.02~eV. The effect is more pronounced in the binuclear and trinuclear complexes, with increases of 0.08~eV (+2.3\%) in $\mathrm{[Fe_2(OtBu)_6]}$ and 0.13~eV (+5.5\%) in $\mathrm{[Fe_3O(Piv)_6(H_2O)_3]^+}$.
The inclusion of methods to treat strong correlation via DFT+$U$ and hybrid functionals complicates the picture somewhat, but the increased stability of high-spin cases remains, as does the improved agreement with the reference methods in most cases.
These shifts are likely to reflect the enhanced variational freedom afforded by spin-dependent NGWFs, which improves the treatment of spin-channel differences and localization in larger, magnetically complex clusters, particularly in the high-spin case.

In most cases, the inclusion of spin dependence brings ONETEP results into closer agreement with the hybrid-functional PBE0 benchmarks. For example, in $\mathrm{[Fe_2(OtBu)_6]}$, the spin-polarized ONETEP gap (3.63~eV) closely approaches the ORCA value of 3.74~eV. Similarly, the trinuclear cluster $\mathrm{[Fe_3O(Piv)_6(H_2O)_3]^+}$ yields 2.49~eV with spin-dependent NGWFs, only 0.07~eV below the hybrid result. These improvements highlight the importance of spin polarization for accurately describing exchange interactions and localized magnetic states in transition metal clusters.

However, the mononuclear complex $\mathrm{[Fe(H_2O)_6]^{3+}}$ deviates from this trend. While the spin-dependent NGWFs still increase the HS-LS splitting, both overestimate its value compared to the ORCA baseline (1.69~eV), in contrast to the improved agreement seen in other systems. This discrepancy can be attributed to several physical and methodological factors. First, $\mathrm{[Fe(H_2O)_6]^{3+}}$ has a highly symmetric octahedral geometry and weak-field ligands, resulting in a small ligand field splitting that makes the relative spin-state energies highly sensitive to the balance of exchange and correlation. Second, semilocal functionals like PBE+$U$ suffer from residual self-interaction error that can overstabilize delocalized high-spin states. While spin-dependent NGWFs introduce greater flexibility in the wavefunction, they may also allow further delocalization of the $3d$ orbitals in the absence of exact exchange, exacerbating the overstabilization of the high-spin configuration. These effects are well documented in the literature, where hybrid functionals such as PBE0 or B3LYP have been shown to yield more reliable spin-state energetics for Fe(III) aquo complexes.~\cite{reiher2001reparameterization}

For additional comparison, Quantum ESPRESSO systematically underestimates the spin-state splitting in all systems relative to ORCA, most significantly in the multinuclear clusters. In $\mathrm{[Fe_2(OtBu)_6]}$, the Quantum ESPRESSO value is lower than ORCA by over 1.1~eV, and in $\mathrm{[Fe_3O(Piv)_6(H_2O)_3]^+}$ by 0.24~eV. These discrepancies likely reflect limitations of the plane-wave pseudopotential framework in accurately describing localized $d$ electrons and the magnetic interactions they mediate, particularly when coupled with a uniform Hubbard $U$ correction that lacks environment-specific sensitivity

Overall, the introduction of spin-polarized NGWFs in ONETEP improves the accuracy of spin-state energetics, especially for larger and more correlated transition metal clusters. While the method does not uniformly reduce deviations across all systems, it demonstrates meaningful gains in describing localized magnetic behavior within a linear-scaling framework.

\subsection{Two-dimensional van der Waals Magnetic Materials}
\label{sec:twodimvdW}

As an exemplar of the transition metal trihalides, CrI$_3$ has emerged as a model system for exploring magnetism in two-dimensional van der Waals materials~\cite{huang_layerdep_CrI3_2017,sivadas_stacking_CrI3_2018,chen_TopologicalSpin_CrI3_2018,jiang_stacking_CrI3_2019}. In its monolayer form, CrI$_3$ exhibits out-of-plane ferromagnetic order arising from localized Cr$^{3+}$ ions with a nominal high-spin $t_{2g}^{3}e_{g}^{0}$ configuration. In bilayer geometries, the interlayer exchange interaction becomes strongly dependent on stacking, with both ferromagnetic and antiferromagnetic alignments observed depending on the lateral displacement of adjacent layers~\cite{sarkar2020electronic,xu_coexisting_CrI3_2022}.

\begin{figure*}[]
    \centering
     \includegraphics[width=\linewidth]{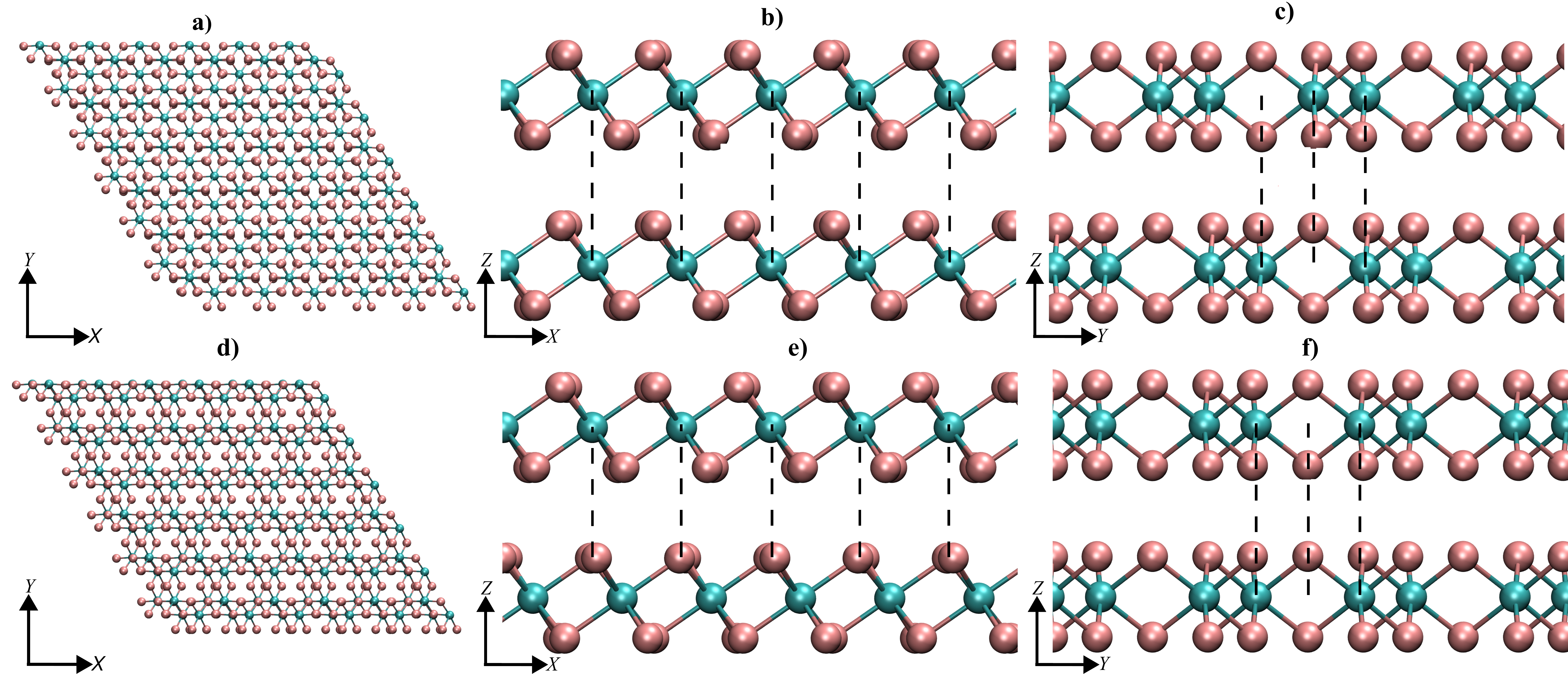}
\caption{Atomic structure of bilayer CrI3 in the low-temperature (top row) and high-temperature (bottom row) phases. Panels (a) and (d) show top views of the bilayer supercells. Panels (b) and (e) present side views in the $XZ$ plane, while panels (c) and (f) show side views in the orthogonal $Y$Z plane.}
  \label{fig:LT_Vs_LT}
\end{figure*}

This stacking dependence is reflected in the existence of two closely related bilayer structures, stabilized at different temperatures, as shown in Fig.~\ref{fig:LT_Vs_LT}. In the low-temperature phase, the bilayers adopt a higher-symmetry stacking in which the local Cr–Cr and Cr–I environments remain equivalent across the bilayer, resulting in a relatively uniform interlayer registry. By contrast, the high-temperature phase is characterized by a symmetry-lowered stacking obtained through a relative in-plane displacement of the layers, which breaks the equivalence of interlayer Cr–Cr alignments and alters the local iodine coordination. Although the intralayer structure of each CrI$_3$ sheet remains largely unchanged, these subtle differences in interlayer geometry have a pronounced impact on the magnetic properties.

In conventional linear-scaling DFT approaches using a shared basis for spin-up and spin-down electrons, the representation of spin polarization is constrained, potentially distorting the energetics of competing magnetic configurations. The spin-dependent NGWF framework, by contrast, allows distinct orbital character and spatial localization for each spin channel, which is particularly critical in systems such as CrI$_3$, where exchange interactions are highly directional and sensitive to intra-atomic spin polarization.

The demonstration of the importance of this enhanced flexibility is part of the reason to study CrI$_3$, while another is its high sensitivity to methodological choices in DFT+$U$ parameterization. In its HT stacking phase, CrI$_3$ exhibits a near-degeneracy between FM and AFM alignments, with energy differences on the order of meV per Cr atom. Jiang~\textit{et al.}\cite{jiang_stacking_CrI3_2019} observed significant dependence of the energy difference $\Delta E_{\mathrm{AFM-FM}} = E_{\mathrm{AFM}}-E_{\mathrm{FM}}$ on the on-site Coulomb $U-J$, with a sign change around $U-J \simeq 1.5$eV. The precise estimation of $\Delta E_{\mathrm{AFM-FM}}$ is therefore a stringent test for any electronic structure method, particularly one based on localized orbitals, and provides an ideal benchmark for evaluating the role of spin-dependent basis sets in capturing the energetics of low-dimensional magnetic systems.

Bilayer CrI$_3$ was investigated in both its high-temperature (HT) and low-temperature (LT) stacking configurations, under ferromagnetic (FM) and antiferromagnetic (AFM) spin orderings. Initial monolayer geometries and cell parameters were obtained from the Computational 2D Materials Database (C2DB)~\cite{C2DB, Sodequist2024C2DBmagnetism} and subsequently relaxed using Quantum ESPRESSO with the OPTB86b-vdW exchange-correlation functional,~\cite{Klimes2011OPTB86b} a 90~Ry plane-wave cutoff, and a $6 \times 6 \times 1$ Monkhorst-Pack k-point mesh. Bilayer models were constructed by stacking and translating the relaxed monolayer, followed by further atomic relaxation at fixed lattice constants, obtaining an interlayer distance of 3.50~\AA{} and 3.48~\AA{} for the LT and HT phases respectively. 

ONETEP calculations were performed on $6 \times 6 \times 1$ supercells derived from the optimized HT and LT geometries, using a kinetic energy cutoff of 1000~eV and NGWF radii of 15~bohr. Both spin-dependent and non-spin-dependent NGWF schemes were tested to evaluate their effect on the relative energetics of FM and AFM states.  Resolving the subtle energy differences between spin configurations required stricter convergence of the density kernel than is typically used in ONETEP ~\cite{Haynes2008_DKopt}.

To assess the impact of intra-atomic exchange interactions, or (depending on one's perspective
on Hubbard-augmented DFT) the 
subspace-averaged static correlation error
on the Cr $3d$ orbitals, 
we perform corrective parameter
calculations including Hund’s $J$.
When used with the rotationally-invariant
DFT+$U$+$J$ functional the Hubbard $U$   primarily corrects self-interaction and promotes orbital localization, while the $J$  introduces an explicit preference for parallel-spin occupancy, stabilizing high-spin configurations. This distinction is potentially particularly important in CrI$_3$, where the competition between interlayer superexchange and intra-atomic Hund’s exchange dictates the magnetic ground state. In our calculations using ONETEP, the $U$ and $J$ parameters are determined  via the 
minimum-tracking linear-response approach based on atomic projectors~\cite{Linscott2018_UJdet, Moynihan2017_scfU}.
For how to calculate $J$ in the more traditional
self-consistent field formulation of linear
response, see Refs.
~\onlinecite{lambert2023use,Moore2024_UJ}.

For comparison purposes, 
approximately equivalent parameters were computed in Quantum ESPRESSO using the long-standing
self-consistent field linear-response formalism of Cococcioni and de Gironcoli, which estimates Hubbard interactions from orbital response to localized potential perturbations.~\cite{Cococcioni2005_LR_U, Timrov2018_LR_UV}. In Quantum Espresso, DFT+$U$ projectors can be chosen either to be of `atomic' form, in which pseudoatomic orbitals corresponding to the pseudopotential are used directly, or 'ortho-atomic', in which case a L\"{o}wdin orthogonalization is applied to the full set of atomic orbitals in the supercell,
while a subset is chosen as Hubbard 
projectors, here Cr $3d$ only.
Either way, there is not direct alignment between ONETEP and QE in terms of either
the projectors (though these are more similar in the QE `atomic' case) or the details
of the linear-response formalism. 

Table~\ref{tab:UJ} summarizes the directly computed Hubbard $U$ and Hund’s $J$ values obtained using linear-response theory in ONETEP and Quantum ESPRESSO. ONETEP yielded $U = 2.97$~eV and $J = 0.37$~eV, while Quantum ESPRESSO produced $U = 2.73$~eV and $J=0.48$~eV when atomic projectors are used, compared to $U = 4.371$~eV and $J = 0.48$~eV with `ortho-atomic' projectors. The former QE result is in fair numerical agreement with ONETEP, whereas the latter is quite different. This indicates the localisation of DFT+$U$ 3$d$ projectors is a key distinction in this case. The `ortho-atomic' projectors orthonormalize the 3$d$ PAOs against the 5$p$ PAOs of the Iodine, resulting in a more localized projector when compared to 'atomic' projectors. This results in higher values of $U$ and $J$. The projectors in ONETEP are much more akin to the standard `atomic' projectors in QE, but on the other hand, the behavior in terms of orbital occupations of the 'ortho-atomic' projectors is more physically realistic, so we retain both sets of results for comparison.

\begin{table}[h]
\centering
\begin{tabular}{ccc}
\hline
Code & $U$ (eV) & $J$ (eV) \\ \hline
ONETEP (conventional) & 2.95  & 0.37 \\
Contribution of   $\hat{v}^{\sigma}_\textrm{PAW}$ to the above & 1.59 & 0.20 \\
ONETEP (constrained) &
1.77 & 0.18 \\
Quantum ESPRESSO (ortho-atomic)    & 4.37 & 0.57 \\
Quantum ESPRESSO (atomic) & 2.73 & 0.48 \\
\hline
\end{tabular}
\caption{Directly computed Hubbard $U$ and Hund's $J$ values, calculated using ONETEP (and minimum-tracking
linear response
in conventional~\cite{Moore2024_UJ} (a.k.a.
`scaled $2 \times 2$' ~\cite{Linscott2018_UJdet} mode), 
with atomic projectors) and Quantum Espresso
(self-consistent field linear response, with scalar inversion) using both ortho-atomic
and atomic projectors.
Provided also are the explicit
contribution made by PAW to the
ONETEP $U$ and $J$ values, as
well as the values when 
effectively constraining $N$ while $M$
is varying with $\beta$ for $J$,
or fixing $M$ while $N$
is varying with $\alpha$ for $U$,
a.k.a.  `simple $2 \times 2$'.
From the latter, of use for
flat-plane based functionals, 
we also report that 
$f^\uparrow  = 2.01$~eV
and
$ f^\downarrow = 1.17$~eV.
}
\label{tab:UJ}
\end{table}
\FloatBarrier

\begin{figure}[h]
    \centering
    \includegraphics[width=\linewidth]{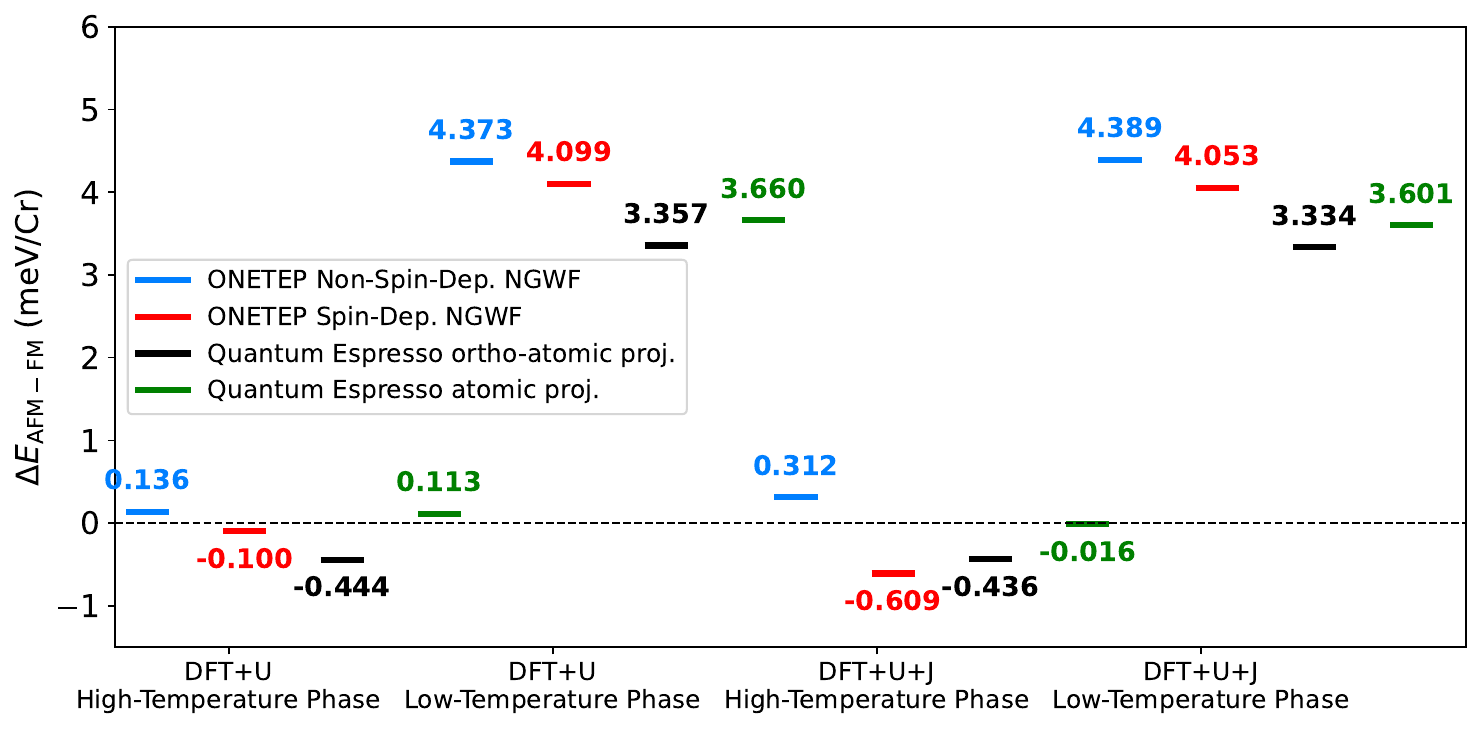}
    \caption{AFM–FM energy differences (in meV/Cr) for CrI$_3$ bilayers in both high-temperature (HT) and low-temperature (LT) stacking configurations. Results are shown for DFT+$U$ and DFT+$U$+$J$, comparing ONETEP with non-spin-dependent NGWFs (blue), ONETEP with spin-dependent NGWFs (red), Quantum Espresso with ortho-atomic projectors (black) and Quantum Espresso with atomic projectors (green).}
    \label{fig:afm_fm_energy_comparison_2}
\end{figure}

Figure~\ref{fig:afm_fm_energy_comparison_2} summarizes the computed AFM-FM energy differences for both the HT and LT bilayer phases under DFT+$U$ and DFT+$U$+$J$, comparing Quantum ESPRESSO (with both projector schemes) to ONETEP (with both non-spin-dependent and spin-dependent NGWFs). 

Some key trends can be seen easily: $\Delta E_{\mathrm{AFM-FM}}$ is much lower for the HT phase, and in most cases is negative, consistent with interlayer AFM ordering being observed at high temperatures. Since $\Delta E_{\mathrm{AFM-FM}}$ is so close to zero, methodological choices can switch it to an unphysical positive value, such as with `atomic' projectors in DFT+$U$. 

The inclusion of $J$ is not particularly influential, though it does in most cases further stabilize the AFM ordering in the HT phase. Notably, for atomic projectors it changes the sign of $\Delta E_{\mathrm{AFM-FM}}$ in the HT phase to restore the correct physics. 

Most notably for our purposes in this work, the calculation using non-spin-dependent NGWFs failed to capture the correct sign of $\Delta E_{\mathrm{AFM-FM}}$, predicting +0.136~meV/Cr and +0.312~meV/Cr, with and without the $J$ correction respectively. By contrast this falls to -0.100~meV/Cr and -0.609~meV/Cr when spin-dependent NGWFs are used. 

In the LT stacking configuration, which favors ferromagnetic alignment, all methods correctly predicted a positive energy difference. With DFT+$U$, ONETEP yielded +4.373~meV/Cr using non-spin-dependent NGWFs and +4.099~meV/Cr with spin-dependent NGWFs, compared to +3.537~meV/Cr from Quantum ESPRESSO. Upon inclusion of Hund’s exchange via DFT+$U$+$J$, the ONETEP values became +4.389~meV/Cr and +4.053~meV/Cr, respectively. Although the absolute values differ slightly, the inclusion of spin-dependent NGWFs consistently improved agreement with the QE benchmark results. These results highlight that variational spin freedom is not only necessary for predicting the correct qualitative ordering in borderline cases, such as the HT phase, but also quantitatively improves total energy estimates, even in well-ordered magnetic phases.

The Hund’s $J$ term explicitly favors parallel-spin configurations on the same atomic site, stabilizing high-spin states and enforcing correct spin alignment within open-shell ions. This is essential in systems such as CrI$_3$, where interlayer magnetic ordering arises from a delicate competition between superexchange and intra-atomic exchange. By combining DFT+$U$+$J$ with spin-dependent NGWFs, the ONETEP framework recovers both orbital localization and spin-selective exchange interactions.

The comparisons indicate that the ortho-atomic projectors in Quantum ESPRESSO may be providing a more realistically-localized representation of the correlated Cr~3$d$ subspace for CrI$_3$ than the atomic projectors. This may lead to better separation in real space the transition-metal $d$ character from the surrounding I~$5p$ manifold. This improved separation results in higher computed values of $U$ and $J$, which enhances the energy difference between antiferromagnetic (AFM) and ferromagnetic (FM) states. As a result, the energy difference, $\Delta E_{\mathrm{AFM-FM}}$, becomes more negative in the high-temperature phase, reinforcing the tendency toward interlayer antiferromagnetism.
 
The remaining discrepancy between ONETEP and Quantum ESPRESSO (with ortho-atomic projectors) in Fig.~\ref{fig:afm_fm_energy_comparison_2} can therefore likely be attributed to the different effective localization of the Hubbard projectors. Within ONETEP, it would be possible to achieve a similar increase in localization by constructing more compact Cr~3$d$ projectors, for instance, by generating pseudoatomic orbitals in a higher oxidation state or by explicitly constraining the projector radius to a smaller range. This adjustment would be expected to bring $U$ and $J$ closer to the ortho-atomic values and to shift $\Delta E_{\mathrm{AFM-FM}}$ in the same direction observed when transitioning from atomic to ortho-atomic projectors in Quantum ESPRESSO, reducing the remaining quantitative differences between the two.

To contextualize our findings, we note that prior first-principles studies have reported FM–AFM energy splittings in bilayer CrI$_3$ on the order of sub-meV to several meV per Cr atom, depending on stacking geometry and the exchange-correlation functional used.~\cite{sivadas_stacking_CrI3_2018} For instance, Sivadas~\emph{et al.} found that the low-temperature stacking favors FM order by 3.2~meV/Cr, while the high-temperature stacking shows a slight AFM preference by about 0.5~meV/Cr. While we do not aim to reproduce these results directly, our spin-dependent NGWF calculations are qualitatively consistent with this trend and quantitatively match the FM–AFM energy differences obtained from our own plane-wave DFT benchmarks using Quantum ESPRESSO. By contrast, the non-spin-dependent NGWF approach often misrepresents the energetic ordering—either overestimating the splitting or predicting the incorrect ground state. These discrepancies reinforce the need for spin-adaptive basis sets in systems where subtle magnetic interactions, such as field- and pressure-tunable ordering, play a crucial role.~\cite{huang_layerdep_CrI3_2017, Li2019NatMater}

\begin{figure}[h]
    \centering
    \begin{subfigure}[]{0.31\columnwidth}
        \raisebox{0.9\height}{\makebox[0pt][l]{\textbf{a)}}}
        \includegraphics[width=\columnwidth]{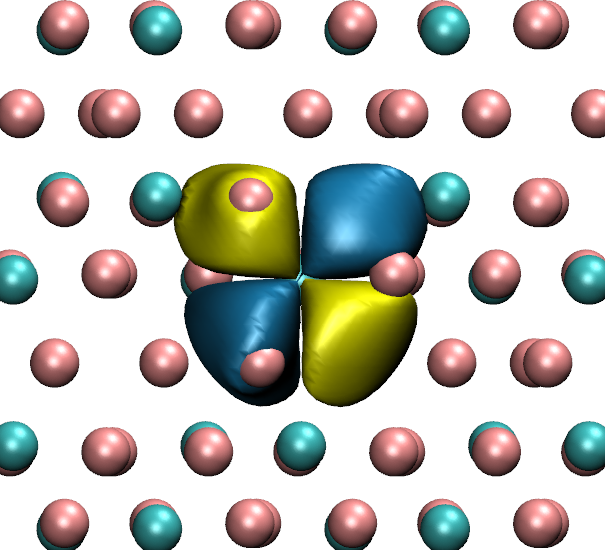}
        \label{fig:ngwf_nsd}
    \end{subfigure}
    \hfill
    \begin{subfigure}[]{0.31\columnwidth}
        \raisebox{0.9\height}{\makebox[0pt][l]{\textbf{b)}}}
        \includegraphics[width=\columnwidth]{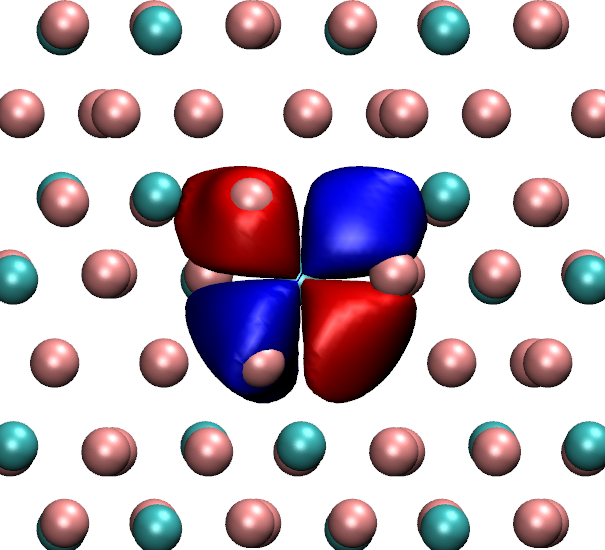}
        \label{fig:ngwf_sd_up_241}
    \end{subfigure}
    \hfill
    \begin{subfigure}[]{0.31\columnwidth}
        \raisebox{0.9\height}{\makebox[0pt][l]{\textbf{c)}}}
        \includegraphics[width=\columnwidth]{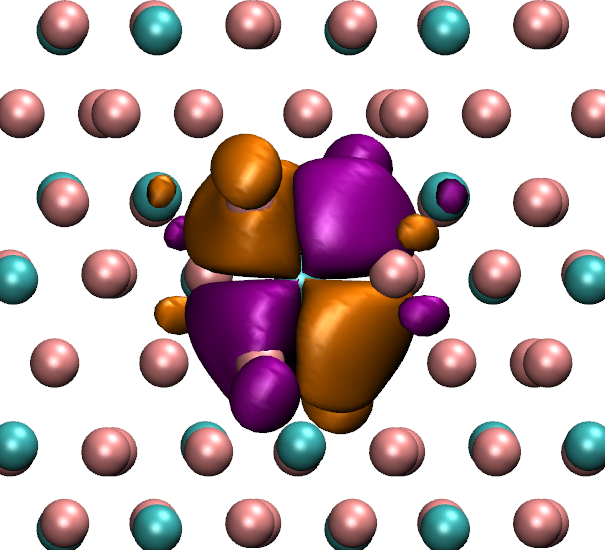}
        \label{fig:ngwf_sd_down_241}
    \end{subfigure}

    \begin{subfigure}[]{0.31\columnwidth}
        \raisebox{0.9\height}{\makebox[0pt][l]{\textbf{d)}}}
        \includegraphics[width=\columnwidth]{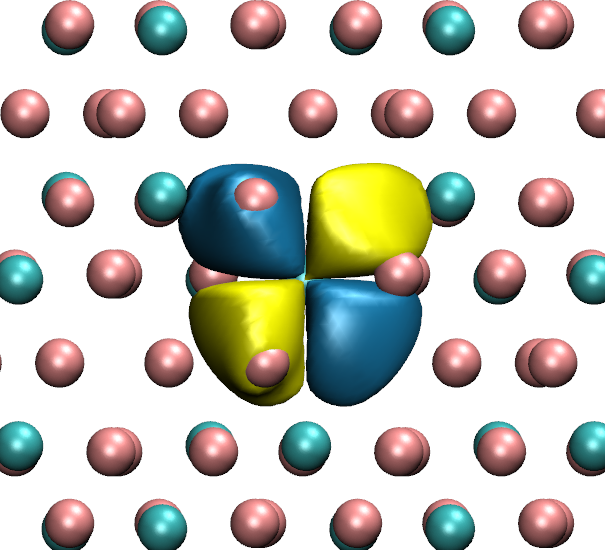}
        \label{fig:ngwf_nsd_235}
    \end{subfigure}
    \hfill
    \begin{subfigure}[]{0.31\columnwidth}
        \raisebox{0.9\height}{\makebox[0pt][l]{\textbf{e)}}}
        \includegraphics[width=\columnwidth]{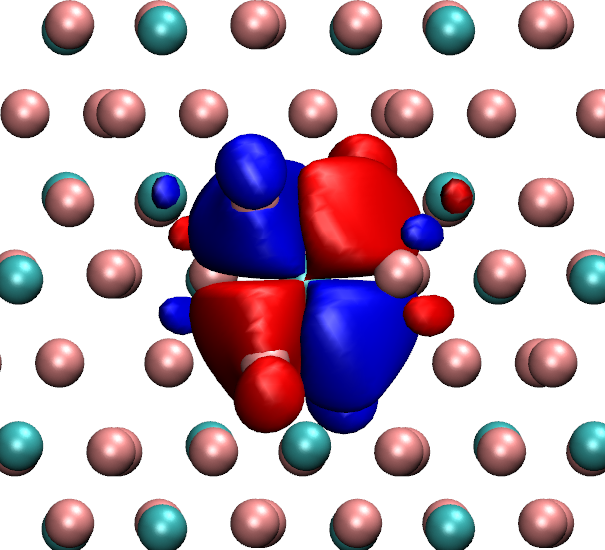}
        \label{fig:ngwf_sd_up_235}
    \end{subfigure}
    \hfill
    \begin{subfigure}[]{0.31\columnwidth}
        \raisebox{0.9\height}{\makebox[0pt][l]{\textbf{f)}}}
        \includegraphics[width=\columnwidth]{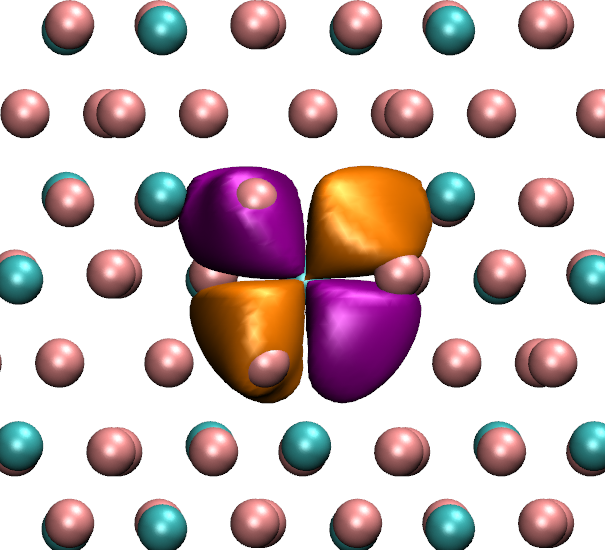}
        \label{fig:ngwf_sd_down_345}
    \end{subfigure}
\caption{Comparison of optimized $d-$orbital NGWFs centered on Cr atoms in bilayer AFM CrI$_3$, demonstrating the effect of spin-dependent variational freedom. Panels a)–c) show NGWFs for a Cr atom in the top layer with net positive spin moment, while panels d)–f) correspond to a Cr atom in the bottom layer with net negative spin moment. In each row, panels a) and d) show the non-spin-dependent NGWFs. Panels b) and e) display the spin-up NGWFs. Finally, panels c) and f) display the spin-down NGWFs.}
    \label{fig:ngwf_comparison}
\end{figure}

To directly illustrate the effect of spin-dependent variational freedom in NGWFs, we visualized representative orbitals from ONETEP calculations on the antiferromagnetic HT-stacked bilayer. Figure~\ref{fig:ngwf_comparison} displays NGWF isosurfaces for two Cr atoms, one located in the top layer with net positive spin, and the other in the bottom layer with net negative spin. For each site, we show the corresponding non-spin-dependent NGWF (panels a and d), as well as the separately optimized spin-up (panels b and e) and spin-down (panels c and f) NGWFs.

\FloatBarrier

In the non-spin-dependent case, the orbital shapes are spatially symmetric and delocalized, reflecting a compromise between the spin-up and spin-down components. This symmetry masks the local exchange broken symmetry and suppresses the spin-polarized character of the Cr 3$d$ states. By contrast, the spin-dependent NGWFs show clear differences between the two spin channels: on the top-layer Cr site (panels b and c), the spin-up NGWF is compact and localized—characteristic of an occupied majority-spin $t_{2g}^{\uparrow}$ orbital, while the spin-down NGWF is more diffuse and extended, consistent with its minority-spin, unoccupied nature. The pattern is reversed on the bottom-layer Cr site, where the spin-down channel becomes localized (panel f), and the spin-up orbital is more delocalized (panel e).

These visualizations provide direct evidence that spin-dependent NGWFs capture spin polarization not only at the level of total energy but also in the spatial structure of localized orbitals, a crucial requirement for modeling magnetic anisotropy and exchange interactions in layered materials.

In summary, bilayer CrI$_3$ provides a stringent benchmark for evaluating the fidelity of spin-resolved electronic structure methods, due to its stacking-sensitive magnetic phases and meV-scale energy splittings. We have shown that spin-dependent NGWFs are important for correctly predicting the magnetic ground state and improving quantitative agreement with plane-wave DFT in both DFT+$U$ and DFT+$U$+$J$ frameworks. Beyond energetics, spatially resolved NGWF visualizations reveal how spin-dependent variational freedom enables the basis to adapt to local spin polarization.

\subsection{Spin-Dependent NGWFs in Bulk and Nanoscale Ferromagnets}
Metallic ferromagnets present a demanding test for any localized basis set, as they combine itinerant and localized electronic behavior across spin channels. In such systems, magnetic order arises not from strong Coulomb localization, as in correlated insulators, but from imbalance in the exchange interaction between partially filled 3$d$ bands. The resulting exchange splitting leads to differences between the spin channels in terms of orbital hybridization and electronic density near the Fermi level, which can affect the energetics enough to influence the net spin polarization. Accurately capturing this effect requires a variationally flexible basis capable of correctly representing the different orbitals of each spin-channel.

To evaluate how spin-dependent NGWFs perform under these conditions, we applied the method to both bulk and nanoscale cobalt. These are examples of itinerant ferromagnets where localized and delocalized spin physics coexist. Nanoclusters represent an important testing ground for linear-scaling DFT,\cite{Corsini_NanoLetters_2015,Ellaby_PlatNano_2018} yet ferromagnetic transition-metal clusters have received relatively little systematic attention. In this section, we investigated orthorhombic bulk cobalt supercells, and Co nanoclusters containing 13, 55, and 147 atoms, probing magnetic energetics and spin-density distributions from the bulk limit down to low-coordination environments.

\subsubsection*{Bulk Cobalt}

We begin by examining bulk cobalt using both spin-dependent and non-spin-dependent NGWFs to assess how the formalism performs in an extended metallic ferromagnet. The primitive Co unit cell was relaxed using Quantum ESPRESSO with a dense $30\times30\times30$ Monkhorst–Pack k-point grid, and the resulting geometry was used to construct a $6\times6\times6$ (864-atom) supercell for ONETEP single-point calculations using both spin-dependent and non-spin-dependent NGWFs.

Table~\ref{tab:Co_bulk} summarizes the impact of spin-channel adaptivity on key electronic properties. The use of spin-dependent NGWFs lowers the total energy by 30.36~meV per atom compared to the non-spin-depenent case, an appreciable stabilization that arises despite nearly identical integrated magnetization. This indicates that the improved basis freedom yields a more accurate real-space description of spin polarization, which is not captured by the net magnetization alone.

\begin{table}[h]
    \centering
    \caption{
    Energy and spin density differences are computed as (spin-dependent) -- (non-spin-dependent). The absolute spin density refers to the integrated modulus of the spin density field, $|\rho_\uparrow - \rho_\downarrow|$.}
    \label{tab:Co_bulk}
    \begin{tabular}{ccc}
        \hline
        Total energy diff. & Net spin diff. & Abs. spin density diff. \\
        \hline
        -26.24~eV & -6.56 $\mu_\mathrm{B}$/cell& 61.47 $\mu_\mathrm{B}$/cell  \\ \hline
    \end{tabular}
\end{table}
\FloatBarrier

\begin{figure}[h]
    \centering
    \includegraphics[width=0.4\paperwidth]{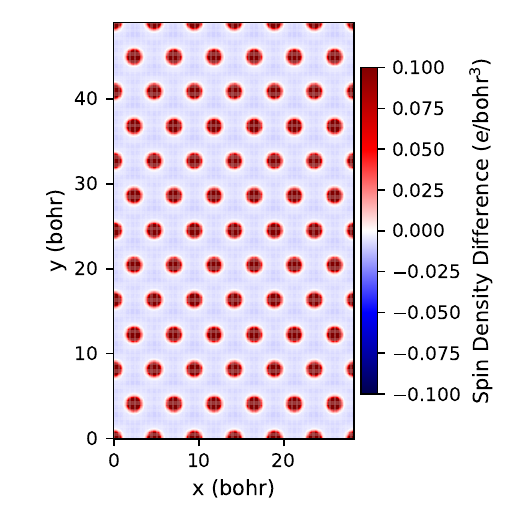}
    \caption{ Spin density difference, $\Delta \rho_s(\mathbf{r})$, shown as a central $z$-slice of the bulk Co supercell.}
    \label{fig:spin_density_diff}
\end{figure}

Figure~\ref{fig:spin_density_diff} provides a spatial view of how spin-dependent NGWFs affect the spin density. 
The plot shows the difference in spin density between the two methods, projected through the central plane of the supercell. 
While the total magnetic moment remains nearly unchanged, the spatial distribution of spin polarization varies substantially: spin-dependent NGWFs recover greater spin localization near Co nuclei and restore the expected spatial decay of minority-spin polarization in interstitial regions. 
For visualization, $\Delta \rho_s(\mathbf{r})$ is shown as a two-dimensional slice using a symmetric color map (red: positive, blue: negative) centered at zero. 

\FloatBarrier

In the non-spin-dependent case, the NGWF basis is constrained to serve both spin channels equally, forcing a spatial averaging of spin-up and spin-down features. This leads to smoother, more delocalized spin densities that fail to capture the distinct features of the two spin channels.
In contrast, spin-dependent NGWFs allow majority and minority spin orbitals to localize independently in space. This freedom may be resulting in a more accurate reconstruction of sharp spin density gradients, both near atomic cores, where exchange interactions are strongest, and in interstitial regions, where itinerant spin polarization emerges. The observed increase of 61.47 units in the integrated absolute spin density, despite minimal change in net spin, signals that the spin-dependent formalism recovers spatial detail in the spin density that is otherwise averaged out.

\begin{figure}[ht]
    \centering
    \includegraphics[width=0.95\columnwidth]{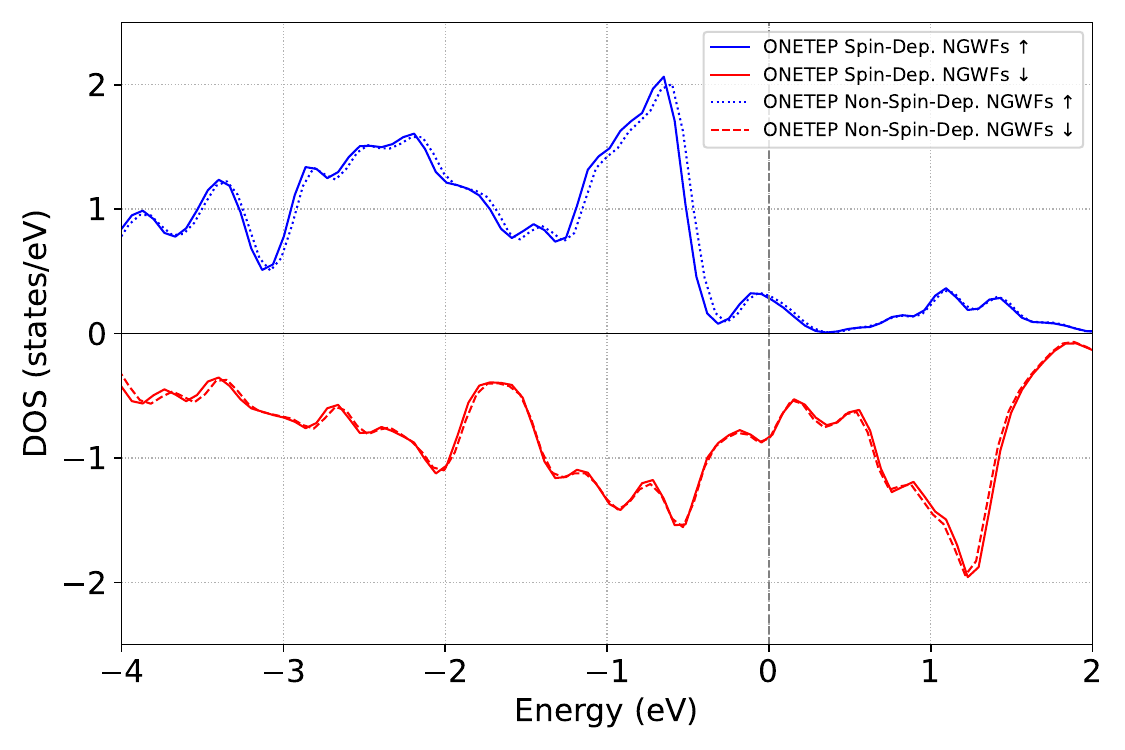}
    \caption{Spin-resolved density of states (DOS) for bulk Co computed using ONETEP with spin-dependent NGWFs (solid) and non-spin-dependent NGWFs (dashed). Spin-up and spin-down channels are shown in blue and red, respectively.}
    \label{fig:Co_DOS}
\end{figure}

To quantify how spin-adaptive basis sets affect the electronic spectrum, we computed the spin-resolved density of states (DOS) for bulk Co using both spin-dependent and non-spin-dependent NGWFs. As shown in Figure~\ref{fig:Co_DOS}, both methods capture the main features of Co's metallic DOS, namely the peaks associated with occupied $3d$ states, a spin-polarized valence manifold, and a finite DOS at the Fermi level. However, the use of spin-dependent NGWFs produces substantial shifts in peak positions even after the Fermi levels of both calculations have been aligned at zero on the scale, particularly noticeable in the exchange-split peaks below and above the Fermi level.

The improved energy fidelity of occupied states carries important energetic consequences: in the spin-dependent case, low-energy $3d$ states, especially in the majority-spin channel, are more localized and filled, reflecting a more efficient variational occupation of the manifold. In contrast, the non-spin-dependent basis exhibits smeared features near the Fermi level due to enforced averaging between spin channels. This aligns with the energy differences reported in Table~\ref{tab:Co_bulk}, suggesting that the energetic gain arises from enhanced spatial flexibility and better representation of the occupancies around the Fermi level.

\subsubsection*{Cobalt Nanoclusters}

\begin{figure}[h]
    \centering
    \begin{subfigure}[]{0.31\columnwidth}
        \raisebox{0.9\height}{\makebox[0pt][l]{\textbf{a)}}}
        \includegraphics[width=\columnwidth]{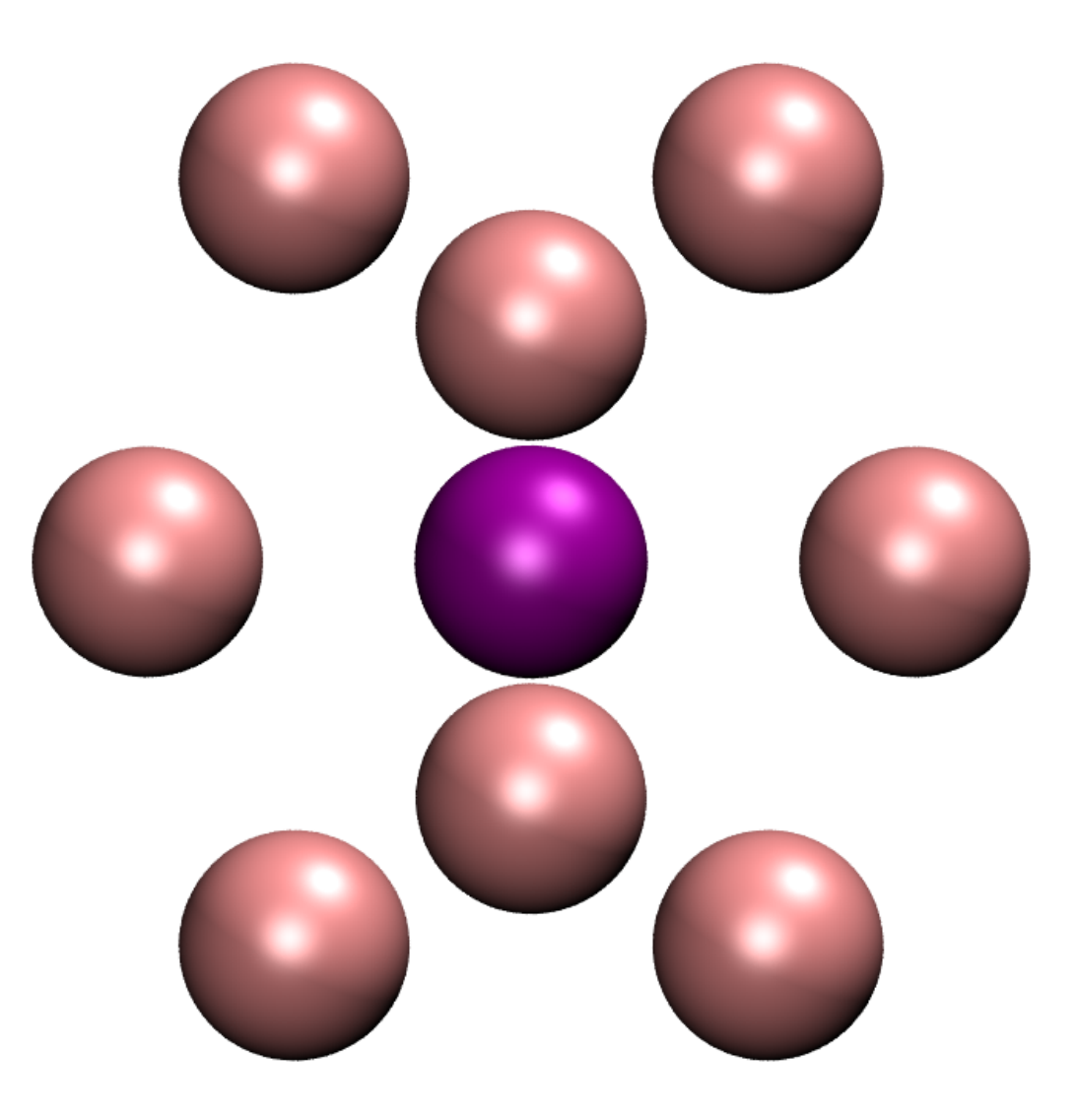}
    \end{subfigure}
    \hfill
    \begin{subfigure}[]{0.31\columnwidth}
        \raisebox{0.9\height}{\makebox[0pt][l]{\textbf{b)}}}
        \includegraphics[width=\columnwidth]{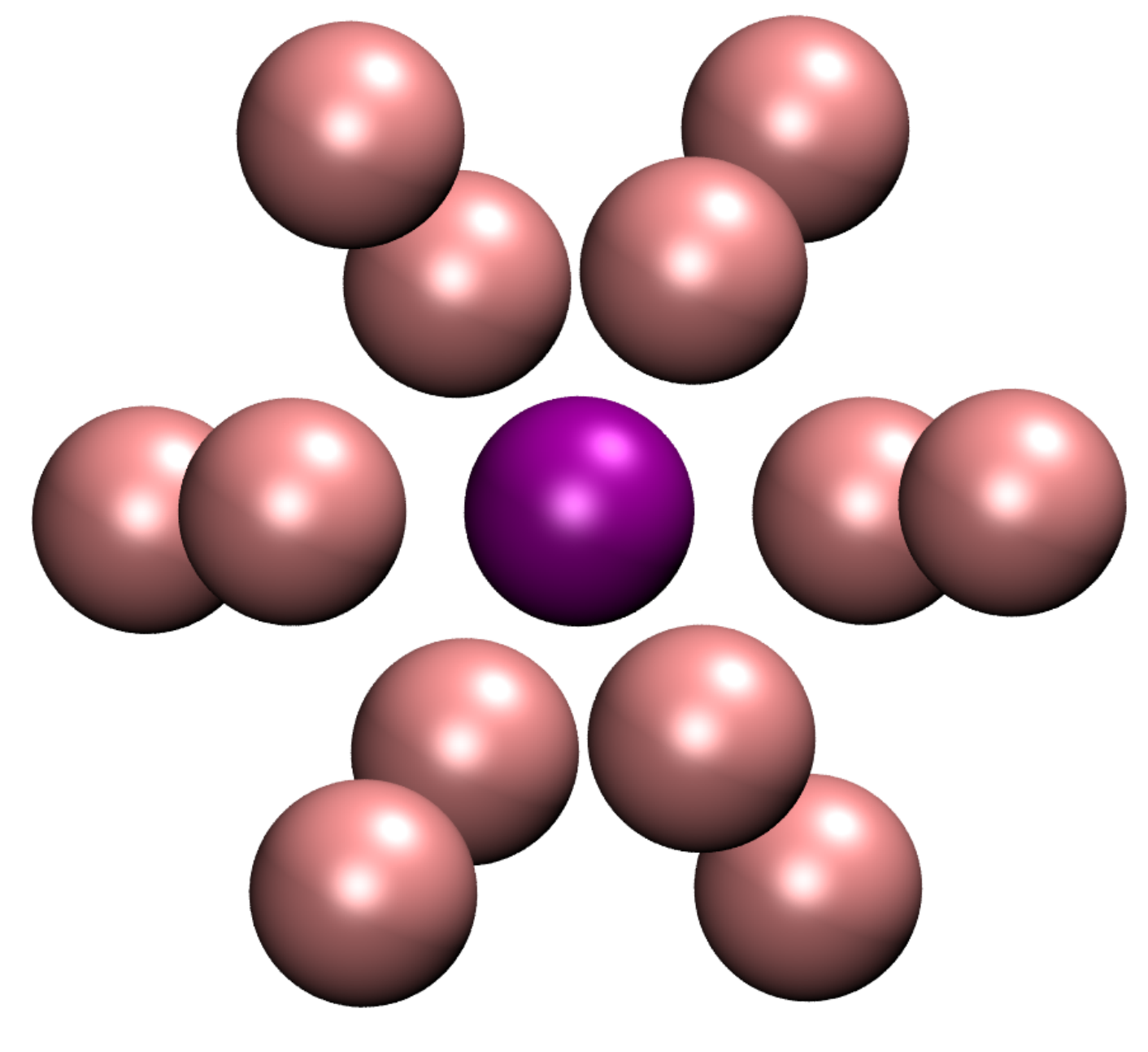}
    \end{subfigure}
    \hfill
    \begin{subfigure}[]{0.31\columnwidth}
        \raisebox{0.9\height}{\makebox[0pt][l]{\textbf{c)}}}
        \includegraphics[width=\columnwidth]{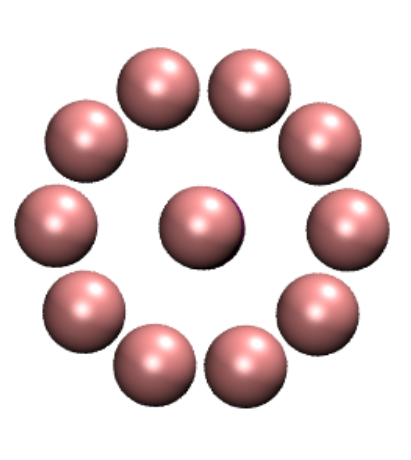}
    \end{subfigure}

    \begin{subfigure}[]{0.31\columnwidth}
        \raisebox{0.9\height}{\makebox[0pt][l]{\textbf{d)}}}
        \includegraphics[width=\columnwidth]{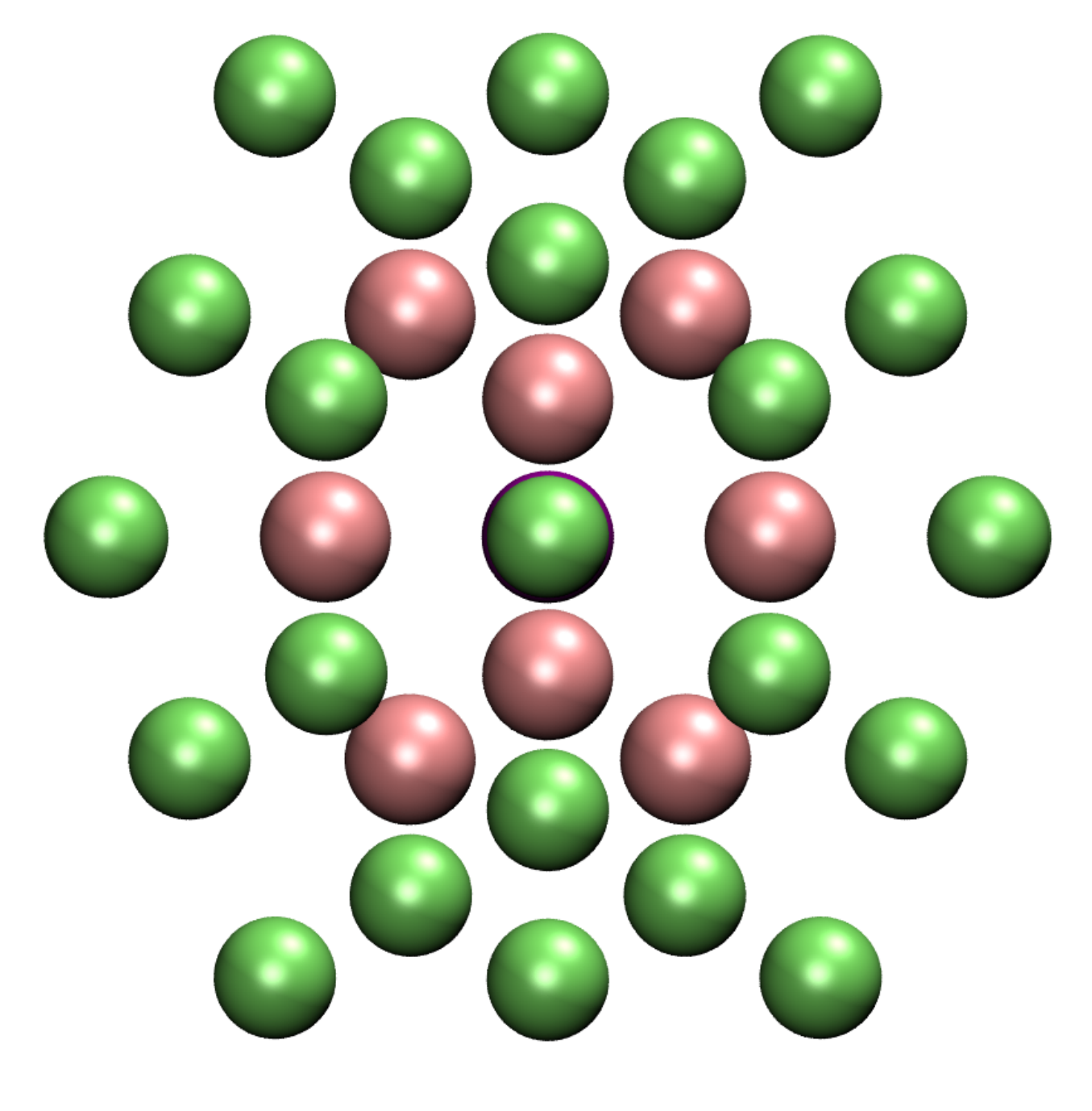}
    \end{subfigure}
    \hfill
    \begin{subfigure}[]{0.31\columnwidth}
        \raisebox{0.9\height}{\makebox[0pt][l]{\textbf{e)}}}
        \includegraphics[width=\columnwidth]{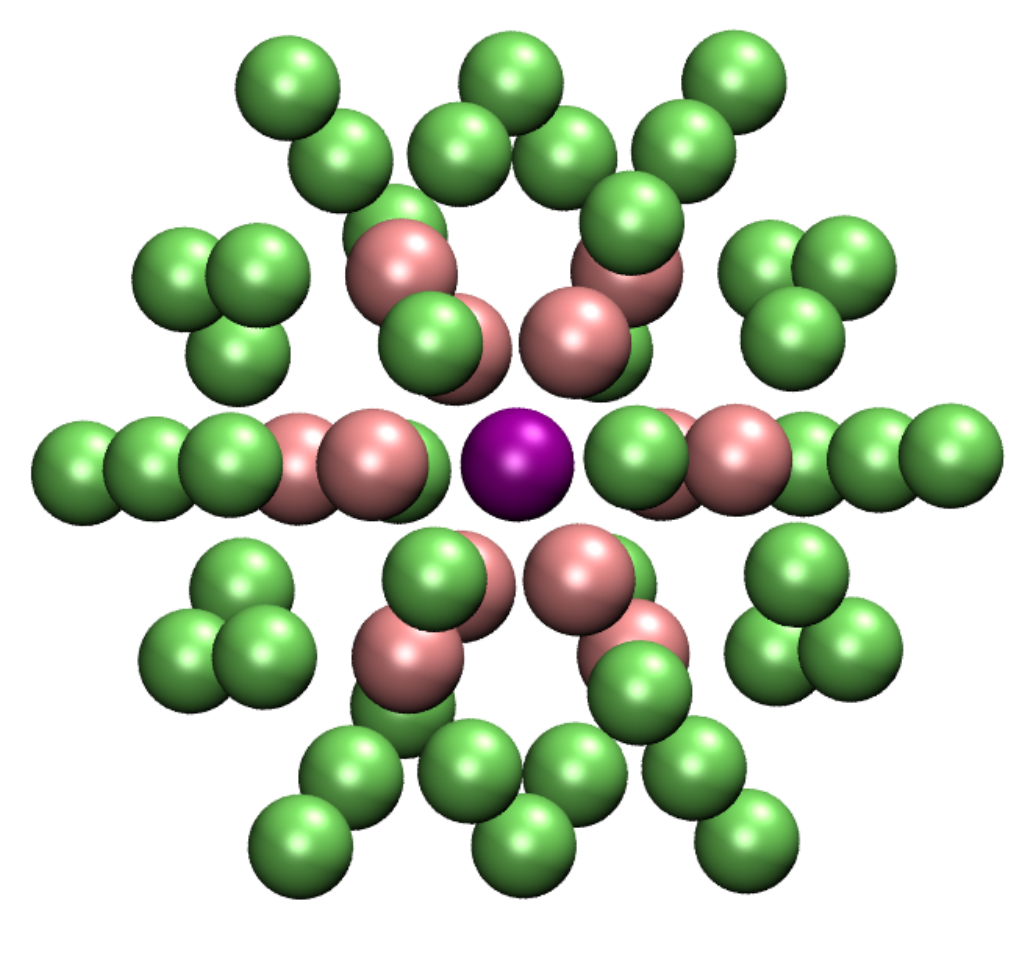}
    \end{subfigure}
    \hfill
    \begin{subfigure}[]{0.31\columnwidth}
        \raisebox{0.9\height}{\makebox[0pt][l]{\textbf{f)}}}
        \includegraphics[width=\columnwidth]{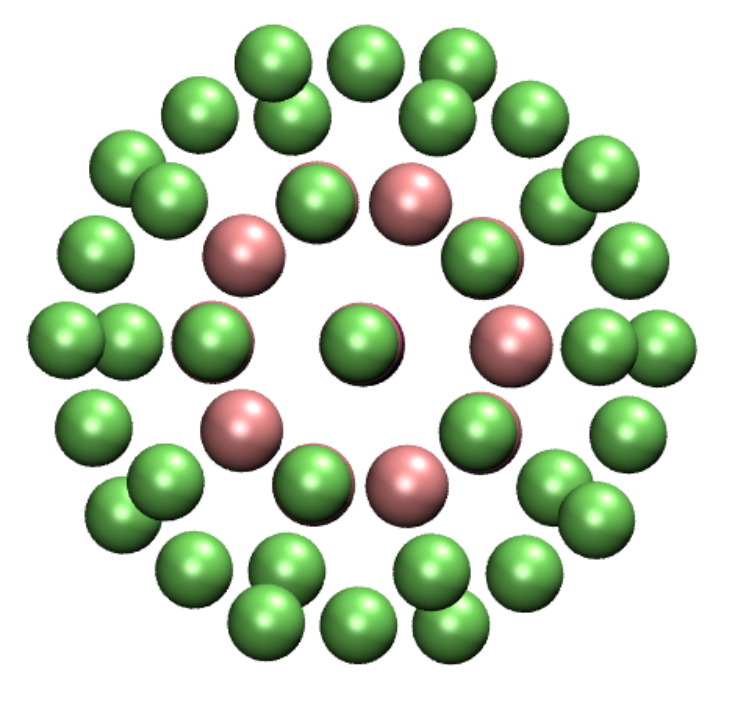}
    \end{subfigure}

    \begin{subfigure}[]{0.31\columnwidth}
        \raisebox{0.9\height}{\makebox[0pt][l]{\textbf{g)}}}
        \includegraphics[width=\columnwidth]{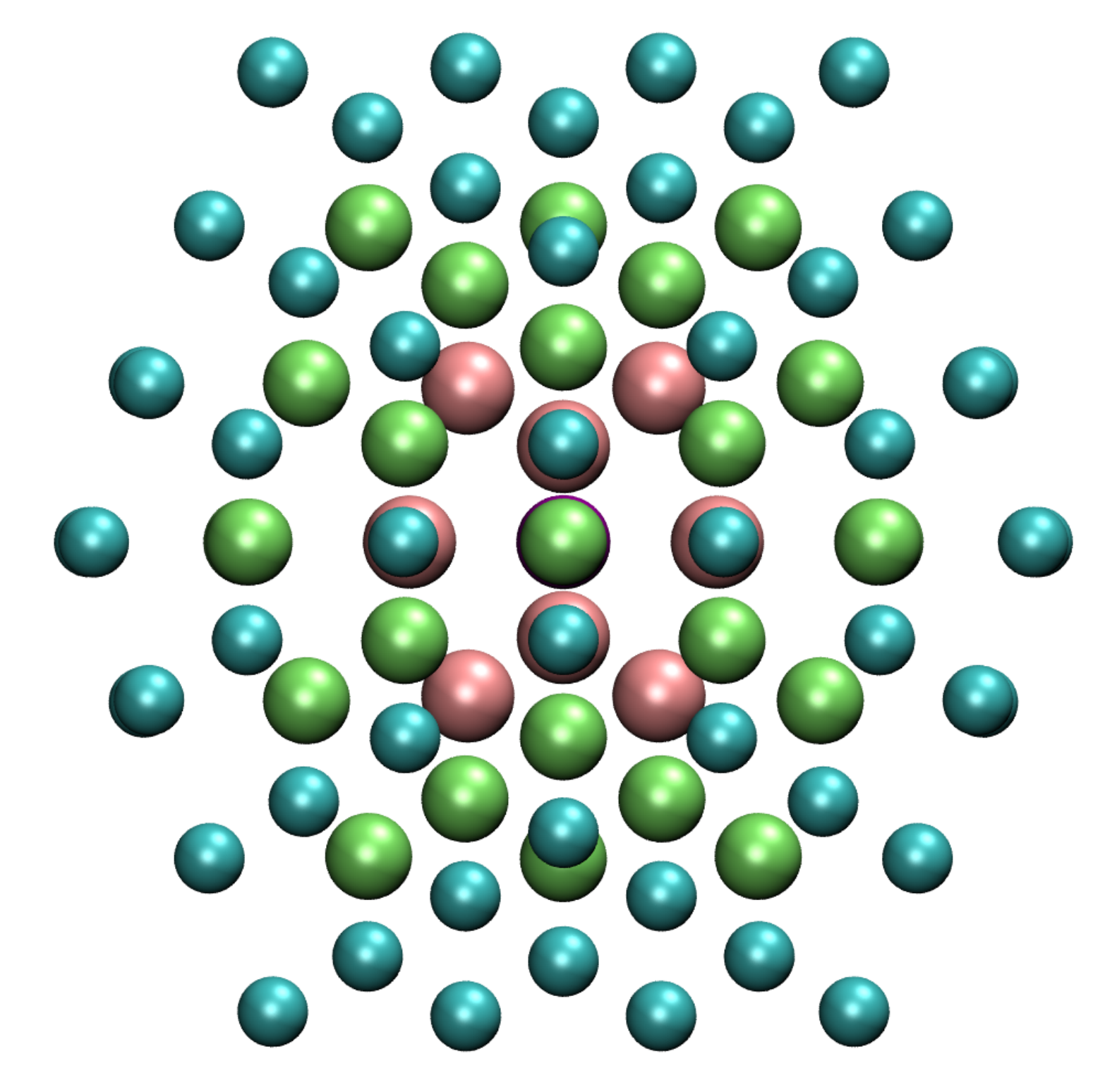}
    \end{subfigure}
    \hfill
    \begin{subfigure}[]{0.31\columnwidth}
        \raisebox{0.9\height}{\makebox[0pt][l]{\textbf{h)}}}
        \includegraphics[width=\columnwidth]{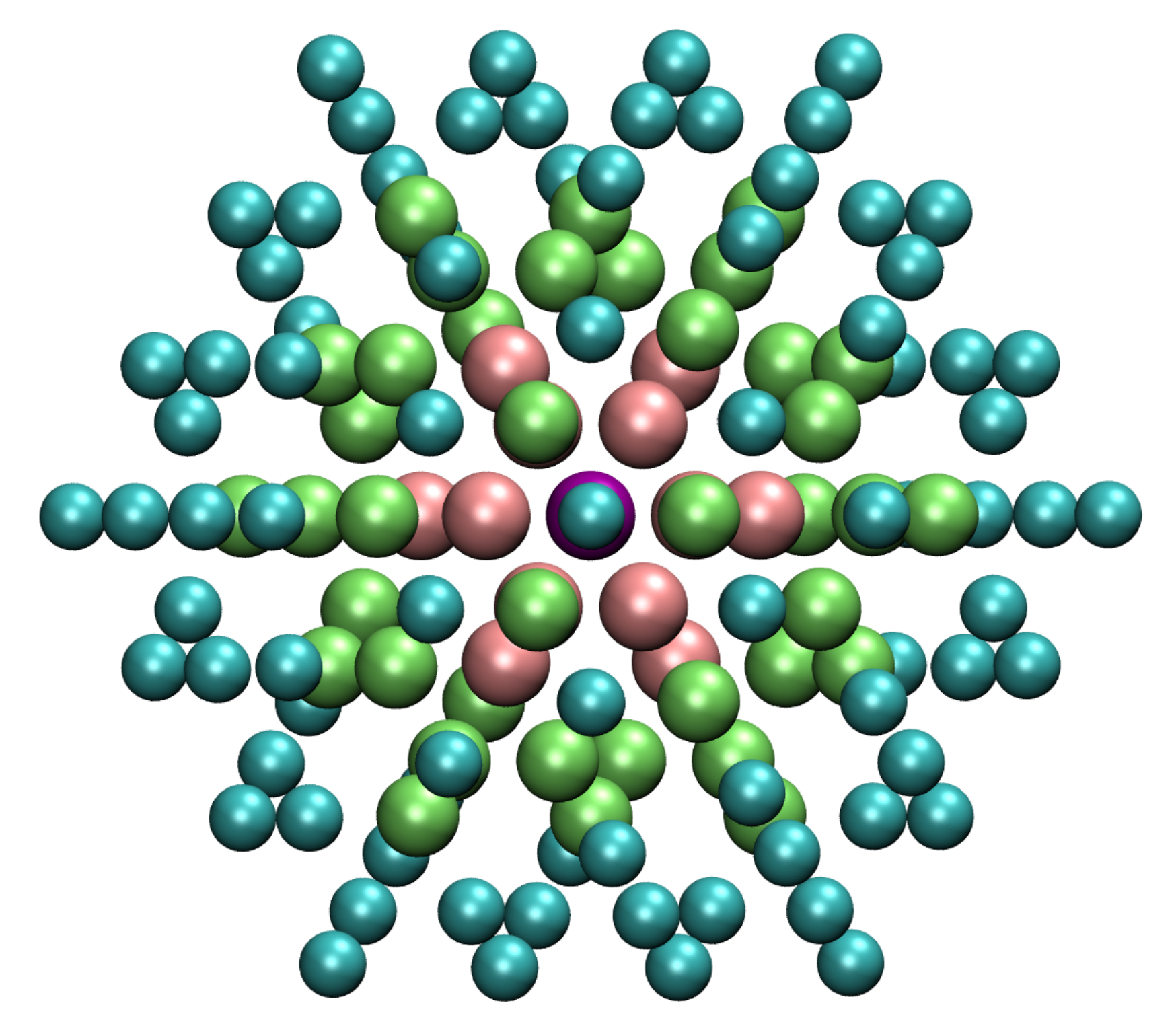}
    \end{subfigure}
    \hfill
    \begin{subfigure}[]{0.31\columnwidth}
        \raisebox{0.9\height}{\makebox[0pt][l]{\textbf{i)}}}
        \includegraphics[width=\columnwidth]{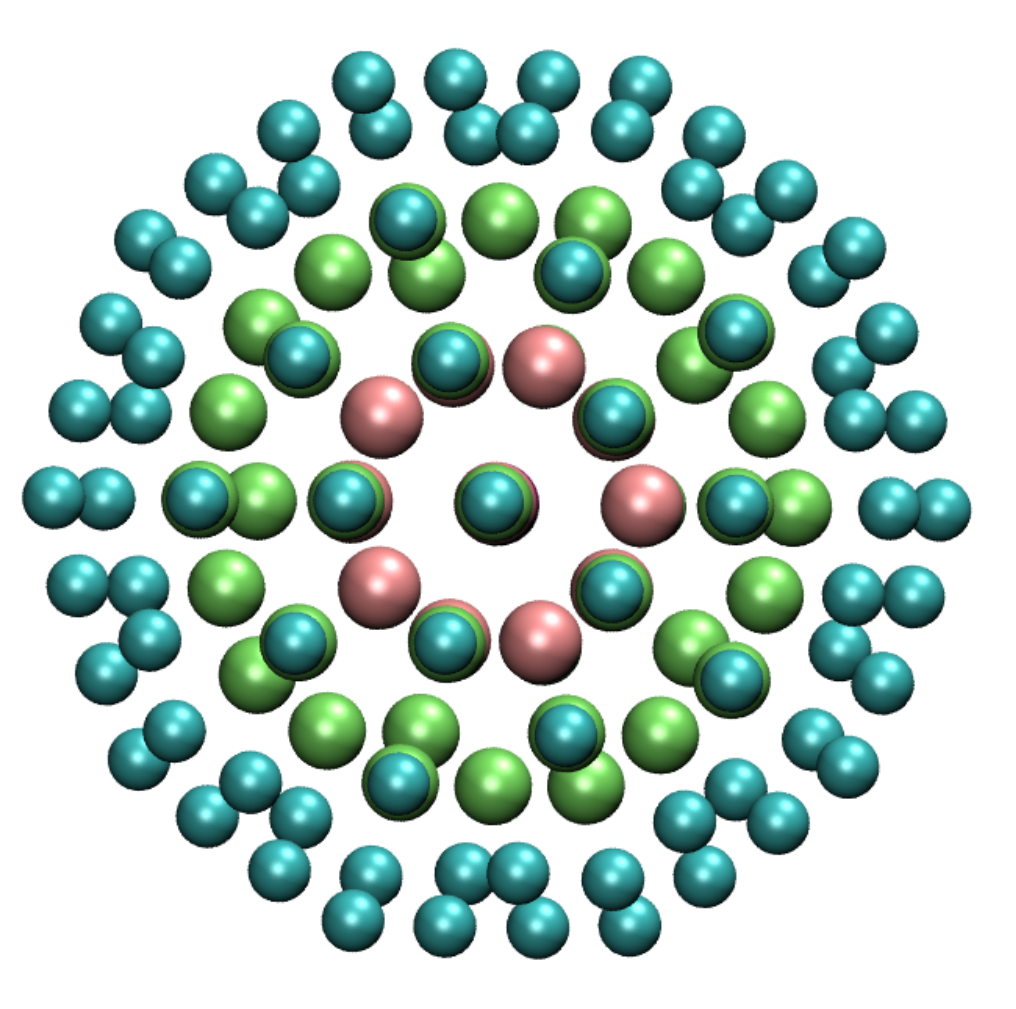}
    \end{subfigure}
\caption{Atomic structures of three Co cluster geometries with increasing size and coordination. Panels (a–c), (d–f), and (g–i) correspond to clusters 1, 2, and 3, respectively, shown in top and oblique views. All atoms are Co, with colours used solely to distinguish successive coordination shells for visualization purposes: the central atom is shown in purple, first-shell atoms in brown, second-shell atoms in green, and third-shell atoms in blue.}
    \label{fig:Co_clusters}
\end{figure}

We next perform calculations on nanoscale Co clusters (Co$_{13}$, Co$_{55}$, and Co$_{147}$) to evaluate spin-dependent NGWFs in finite systems with varying coordination environments. Each structure was fully relaxed under spin-polarized conditions, using both spin-dependent and non-spin-dependent NGWFs. The geometries of the Co$_{13}$, Co$_{55}$, and Co$_{147}$ clusters considered in this work are shown in Fig.~\ref{fig:Co_clusters}. These clusters represent successive closed-shell motifs with increasing coordination, providing a systematic progression from surface-dominated to more bulk-like local environments. To aid visualization and comparison between cluster sizes, atoms are grouped into coordination shells relative to the central atom, highlighting the evolution of local bonding environments as the cluster size increases. Both top and oblique views are shown to convey the three-dimensional structure of each cluster.

To probe the role of basis set flexibility, we employed two configurations of the valence NGWF set. The first, denoted $N_\phi=13$, included the 4$p$ orbital alongside the 3$d$ and 4$s$ states, resulting in 13 NGWFs per atom. The second, $N_\phi=10$, excluded the three 4$p$ orbitals, limiting the basis to 10 NGWFs per atom. This comparison allows us to isolate how spin adaptivity and orbital flexibility independently contribute to the electronic and magnetic properties. The contrast between the $N_\phi=13$ and $N_\phi=10$ configurations highlights the connection between orbital flexibility and spin-channel localization. For $N_\phi=13$, the inclusion of 4$p$ orbitals introduces additional variational freedom that can partially offset the limitations of a shared orbital basis between spin channels. However, it also facilitates spatial delocalization of spin density, which can reduce the contrast between majority- and minority-spin states. This leads to smoother spin distributions and slightly reduced energetic gains from using spin-dependent NGWFs. Conversely, the more compact $N_\phi=10$ basis restricts orbital delocalization, forcing spin polarization to concentrate in the 3$d$ manifold. As a result, the benefit of channel-specific basis optimization is amplified. This trade-off illustrates a general principle in spin-polarized DFT: orbital delocalization can weaken the resolution of exchange splitting unless the variational basis is permitted to adapt independently for each spin channel.

Table~\ref{tab:Co_clusters} summarizes the total and per-atom energy differences between spin-dependent and non-spin-depenent NGWF calculations for each cluster and configuration. In all cases, spin-dependent NGWFs yield lower total energies, confirming the advantage of allowing spin channels to optimize independently. The energetic stabilization is more pronounced on a per-atom basis in smaller clusters, consistent with enhanced surface magnetism. Notably, the $N_\phi=10$ configuration exhibits systematically larger energy gains per atom than $N_\phi=13$, emphasizing that when orbital delocalization is suppressed, spin-channel flexibility becomes especially important. These energetic trends correlate strongly with changes in total spin polarization.

\begin{table}[h]
\centering
\caption{Total and per-atom energy differences in eV, using total energies from spin-dependent and non-spin-dependent NGWF calculations, for Co clusters of varying size, and using two different NGWF counts ($N_\phi=13$ and $N_\phi=10$).}
\label{tab:Co_clusters}
\begin{tabular}{ccc}
\hline
\multicolumn{3}{c}{$N_\phi=13$} \\
\hline
\# Co Atoms & Energy Diff.& Energy Diff. per Co \\
\hline
13 & -0.5718 & -0.04398 \\
55 & -2.1767 & -0.03958 \\
147 & -5.2398 & -0.03564 \\
\hline
\multicolumn{3}{c}{$N_\phi=10$ Configuration} \\
\hline
\# Co Atoms & Energy Diff.& Energy Diff. per Co\\
\hline
13 & -0.9290 & -0.07146 \\
55 & -4.8188 & -0.08762 \\
147 & -11.2009 & -0.07620 \\
\hline
\end{tabular}
\end{table}

Integrated spin densities provide further evidence of the benefits of spin-dependent NGWF optimization in cobalt nanoclusters. For each system studied, the total spin polarization was computed using both spin-dependent and non-spin-dependent NGWFs and compared to plane-wave DFT benchmarks from Quantum Espresso where available. These results are collected in Table~\ref{tab:Co_clusters_spin}. For Co$_{147}$, the plane-wave reference is omitted, as a fully converged calculation at this scale was impractical within the scope of this study.

\begin{table*}[ht]
\centering
\caption{Integrated spin densities, in $\mu_\mathrm{B}$, for Co nanoclusters using spin-dependent and non-spin-dependent NGWFs, compared with plane-wave reference values from Quantum ESPRESSO.}
\label{tab:Co_clusters_spin}
\begin{tabular}{cccc}
\hline
\multicolumn{4}{c}{$N_\phi=13$} \\
\hline
\# Co Atoms &non-spin-dependent NGWFs &spin-dependent NGWFs & Quantum ESPRESSO \\
\hline
13 & 27.65 & 28.59 & 30.62 \\
55 & 101.84 & 102.57 & 103.63 \\
147 & 254.26 & 256.02 & - \\
\hline
\multicolumn{4}{c}{$N_\phi=10$} \\
\hline
\# Co Atoms & non-spin-dependent NGWFs & spin-dependent NGWFs & Quantum ESPRESSO \\
\hline
13 & 27.61 & 29.46 & 30.62 \\
55 & 101.80 & 103.48 & 103.63 \\
147 & 250.70 & 257.47 & - \\
\hline
\end{tabular}
\end{table*}

As shown in Table~\ref{tab:Co_clusters_spin}, spin-dependent NGWFs consistently yield higher integrated spin moments than their non-spin-dependent counterparts across all cluster sizes. This presumably reflects greater variational freedom leading to better spin-adaptation and greater local spin polarization. The improvement is most pronounced in the $N_\phi = 10$ calculations, where the absence of 4$p$ orbitals forces the basis to localize more tightly around the 3$d$ manifold, amplifying the benefit of spin dependence, particularly in Co$_{147}$, which shows an increase of nearly 7~$\mu\mathrm{B}$. Agreement with Quantum ESPRESSO is also improved by the spin-dependent formalism, particularly in smaller clusters where localized spin polarization is stronger and easier to capture. 

\begin{figure*}[ht]
    \centering
    \begin{subfigure}[h]{0.45\linewidth}
        \centering
        \includegraphics[width=\columnwidth]{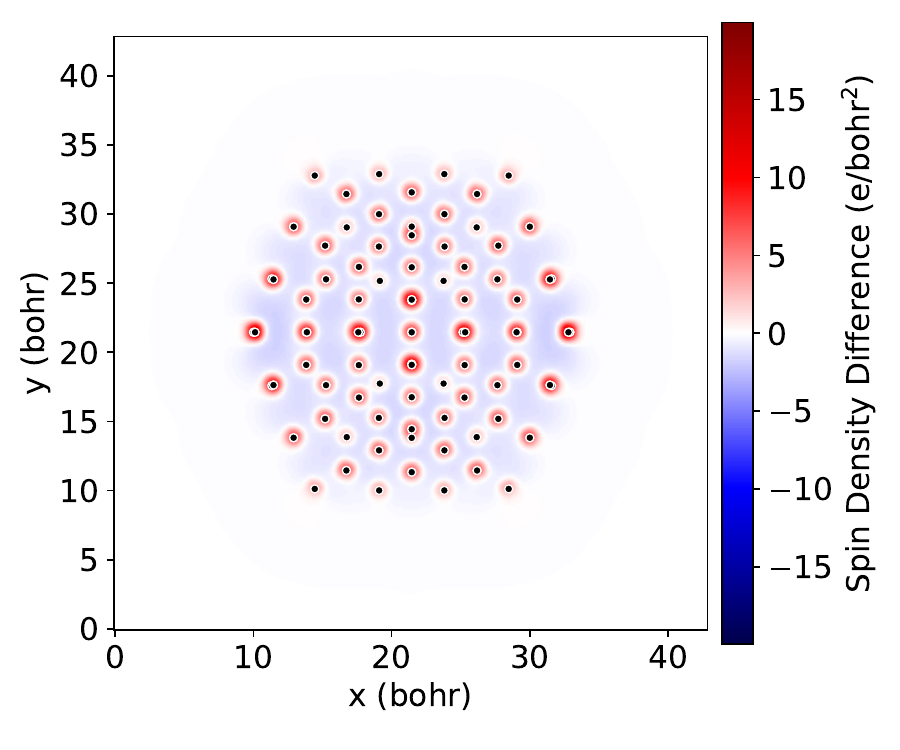}
        \caption{$N_\phi=13$}
    \end{subfigure}
    \hfill
    \begin{subfigure}[h]{0.45\linewidth}
        \centering
        \includegraphics[width=\columnwidth]{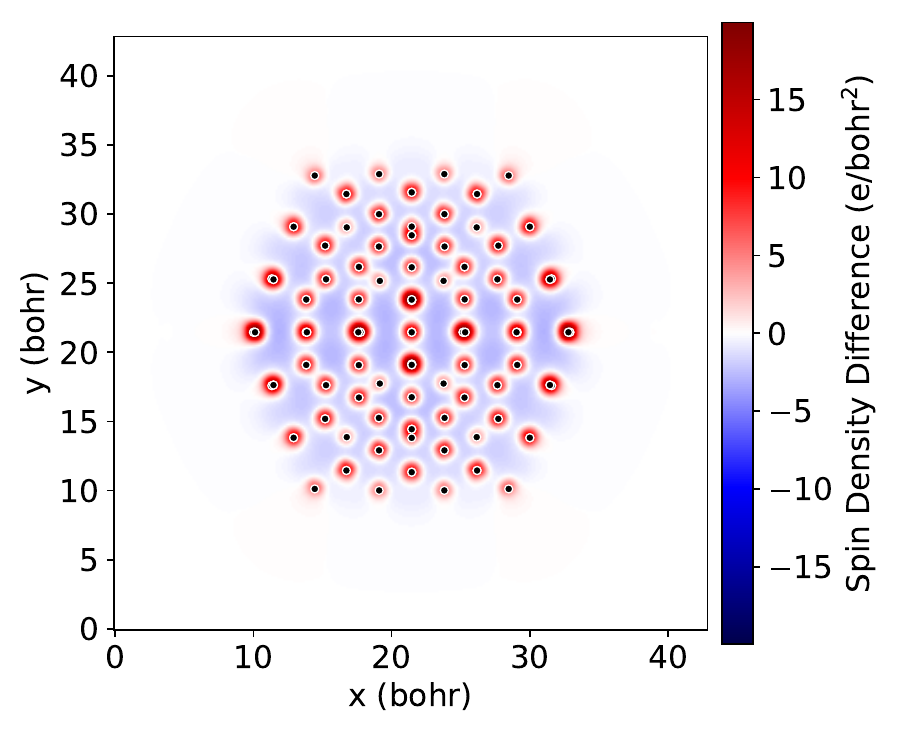}
        \caption{$N_\phi=10$}
    \end{subfigure}
    \caption{
    Projected spin density difference $\Delta \rho_s(\mathbf{r}) = \rho_s^{\mathrm{spin-dependent}}(\mathbf{r}) - \rho_s^{\mathrm{non-spin-dependent}}(\mathbf{r})$ for a given NGWF representation for Co$_{147}$. The black dots denote the position of the Co atoms. Colorbars are normalized across both panels.
    }
    \label{fig:dspin_projection}
\end{figure*}

To visualize these effects directly, Figure~\ref{fig:dspin_projection} compares the spatial distribution of spin-density differences, between spin-dependent and non-spin-dependent calculations for the $N_\phi=13$ and $N_\phi=10$ basis. The difference field $\Delta \rho_s$ was computed in three dimensions and then projected into the $xy$ plane. For more detail, see Eqs.~\ref{eq:spin_den}-~\ref{eq:integrated_spin_den_diff}.

In both configurations, the enhancement in $\Delta \rho_s$ is concentrated at atomic sites, with the strongest amplitudes appearing near the surface, consistent with reduced coordination and enhanced local magnetic moments. The $N_\phi=10$ configuration, in particular, exhibits stronger radial contrast and greater spatial anisotropy, highlighting that reduced basis flexibility intensifies the benefits of spin-channel separation. These stronger $\Delta \rho_s$ features correlate with larger energy gains from spin-adaptive basis sets (Table~\ref{tab:Co_clusters}) and reflect a more faithful variational description of intra-atomic exchange and magnetic interactions.

These results are consistent with the known size dependence of magnetism in cobalt nanoclusters. As the number of atoms increases, the average spin polarization per atom decreases slightly, reflecting the diminishing influence of surface atoms, which exhibit stronger local moments due to reduced $d$-band hybridization.~\cite{billas1994magnetism,eone2019unraveling} While core atoms become more bulk-like, surface atoms remain magnetically enhanced. Spin-dependent NGWFs capture this transition more accurately than non-spin-depenent ones, particularly when the basis is restricted to localized $3d$ orbitals.

Together, the findings from bulk and nanoscale Co systems demonstrate that spin-dependent NGWFs recover essential features of both localized and itinerant magnetism, improving the accuracy of magnetic properties across dimensionalities. This makes them a powerful tool for scalable, spin-resolved DFT simulations of magnetic materials and nanostructures.

\section{Conclusion and outlook}
We have developed and validated an extension of the nonorthogonal generalized Wannier function formalism within the ONETEP linear-scaling DFT code to support spin-dependent NGWFs. This advancement lifts the constraint of a shared basis between spin channels, enabling the independent optimization of localized orbitals for spin-up and spin-down channels of the density matrix. The resulting formulation restores full variational freedom in spin-polarized systems, equivalent to plane-wave DFT, while preserving ONETEP’s efficiency and scalability.

The implementation is fully compatible with projector augmented-wave (PAW) potentials as well as with DFT+$U$ and DFT+$U$+$J$ methodologies, whose combined use in an LS-DFT context we have detailed here for the first time. This combination allows accurate treatment of exchange and correlation effects in systems with strong local moments. Across a broad suite of benchmark systems, including localized magnetic defects in hBN, high-spin transition-metal complexes, stacked bilayer CrI$_3$, and metallic Co in bulk and nanocluster form, we consistently observe improved total energies, enhanced spin localization, and more accurate prediction of spin-resolved electronic structure.

Notably, in systems where spin polarization plays a central role, such as low-dimensional magnets, transition-metal clusters, and itinerant ferromagnets, the spin-dependent NGWF formalism captures intra-atomic exchange, magnetic anisotropy, and spin-state energetics with accuracy approaching that of plane-wave and hybrid-functional benchmarks.

This work extends the capabilities of linear-scaling DFT to a wider class of spin-polarized materials and provides a foundation for future developments. Integration of spin-dependent NGWFs into hybrid-functional and time-dependent frameworks will further expand the reach of large-scale, accurate \textit{ab initio} simulations of quantum materials, with direct applications in magnetism, spintronics, and correlated electron systems.

The implementation of spin-dependent NGWFs opens multiple avenues for extending the capabilities of linear-scaling DFT, especially in the context of complex magnetic phenomena and correlated materials.

One natural next step is the incorporation of spin–orbit coupling (SOC) within the spin-dependent NGWF framework. SOC plays a central role in a wide range of phenomena, including magnetic anisotropy, topological phases, and Dresselhaus effects, and is essential for the accurate modeling of heavy-element systems and spintronic devices. The variational flexibility afforded by spin-dependent NGWFs provides an ideal foundation for treating SOC within a non-collinear spin formalism, where the NGWFs can accommodate spatially varying spin textures and spin-orbit-induced mixing of spin states.~\cite{gong2019two,tokura2019magnetic,magorrian2024strain}

A promising direction is the use of spin-adaptive support functions combined with machine-learned or physically motivated Hubbard projectors to improve the DFT+$U$ parametrization in low-symmetry or anisotropic magnetic environments. This approach could enhance accuracy in strongly correlated systems beyond what is achievable with standard projector methods.~\cite{yu2020machine,cai2024predicting,uhrin2025machine}

More broadly, the spin-dependent NGWF formalism enhances ONETEP to tackle emerging quantum materials with increasing fidelity, from layered magnets and 2D heterostructures to spin-selective catalysts and molecular spin qubits, while preserving scalability for thousands of atoms. Continued development along these lines will establish ONETEP as a uniquely capable tool for predictive simulations at the frontier of magnetic and correlated materials research.

\section{Data Availability}
Input and output files and processing scripts for all calculations described in this work are available
on GitHub at \href{https://github.com/nickhine/spin_dep_ngwfs}{github.com/nickhine/spin\_dep\_ngwfs}.

\section{Acknowledgements}
The authors gratefully acknowledge support from the EPSRC through the Software for Research Communities grants EP/W029545/1 and EP/W029510/1. AS and NDMH acknowledge support from the Leverhulme Trust Research Project Grant RPG-2022-321. This publication has emanated from research conducted with the financial support of Taighde Éireann — Research Ireland, Grant Number 12/RC/2278\_2, and is co-funded under the European Regional Development Fund under the AMBER award. 
Computational resources were provided by the ARCHER2 UK National Supercomputing Service via the UK Car-Parrinello Consortium (EPSRC Grant EP/X035891/1), the Sulis Tier-2 HPC facility, the Avon cluster at the University of Warwick, the Scientific Computing Research Technology Platform (SCRTP) at the University of Warwick, and the Tryton+ HPC facility at the TASK Supercomputing Centre in Gdańsk, Poland. Sulis is funded by EPSRC Grant EP/T022108/1 and the HPC Midlands+ consortium. These resources were essential for the large-scale simulations presented in this work. We acknowledge Joseph Prentice for his valuable code review efforts and Joseph Prentice, Andrea Greco and José María Escartín Esteban for important preparatory work, which significantly facilitated the implementation of the spin-dependent NGWF formalism.

\bibliography{spin_dep_ngwfs}

\end{document}